\documentclass[conference]{IEEEtran}
\IEEEoverridecommandlockouts
\usepackage{cite}
\usepackage{amsmath,amssymb,amsfonts}
\usepackage{algorithmic}
\usepackage{graphicx}
\usepackage{textcomp}
\usepackage{xcolor}
\def\BibTeX{{\rm B\kern-.05em{\sc i\kern-.025em b}\kern-.08em
    T\kern-.1667em\lower.7ex\hbox{E}\kern-.125emX}}
    
\usepackage{newfloat}
\DeclareFloatingEnvironment[name=TABLE]{tablefig}

\usepackage{url}

\usepackage{float}

\usepackage{hyperref}

\usepackage{hyphenat}
\usepackage{xspace}

\usepackage{amsthm}
\theoremstyle{plain}

\newcommand{\hide}[1]{}

\newcommand{\xhdr}[1]{\vspace{1.7mm}\noindent{{\bf #1.}}}

\newcommand{\ie}{{i.e.}\xspace}

\newcommand{\Secref}[1]{Sec.~\ref{#1}}

\newcommand{\dblsecref}[2]{Sec.~\ref{#1} and \ref{#2}}

\newcommand{\Figref}[1]{Fig.~\ref{#1}}

\newcommand{\Appref}[1]{Appendix~\ref{#1}}

\usepackage{bbm} 
\usepackage{bm}
\usepackage{mathtools}
\DeclarePairedDelimiter\abs{\lvert}{\rvert}
\DeclarePairedDelimiter\norm{\lVert}{\rVert}

\usepackage{amsmath}

\DeclareMathOperator{\adv}{Adv}

\usepackage[normalem]{ulem} 

\DeclareMathOperator{\var}{Var}

\newcommand{\tr}{\text{tr}}
\newcommand{\party}{\text{p}}

\usepackage{centernot}

\usepackage{cryptocode}
\usepackage{wrapfig}

\newcommand\restr[2]{{
  \left.\kern-\nulldelimiterspace 
  #1 
  \right|_{#2} 
  }}

\usepackage{csvsimple}
\usepackage{booktabs,subcaption}

\makeatletter
\newcommand{\linebreakand}{%
  \end{@IEEEauthorhalign}
  \hfill\mbox{}\par
  \mbox{}\hfill\begin{@IEEEauthorhalign}
}
\makeatother

\begin{document}

\title{Distribution inference risks: Identifying and mitigating sources of leakage
\thanks{Valentin Hartmann is supported by Microsoft (Swiss Joint Research Center).}
}

\author{
\IEEEauthorblockN{Valentin Hartmann\IEEEauthorrefmark{1}}
\IEEEcompsocitemizethanks{\IEEEcompsocthanksitem\IEEEauthorrefmark{1}Both authors contributed equally.}
\IEEEauthorblockA{\textit{EPFL} \\
valentin.hartmann@epfl.ch}
\and
\IEEEauthorblockN{Léo Meynent\IEEEauthorrefmark{1}}
\IEEEauthorblockA{\textit{EPFL} \\
leo.meynent@epfl.ch}
\and
\IEEEauthorblockN{Maxime Peyrard}
\IEEEauthorblockA{\textit{EPFL} \\
maxime.peyrard@epfl.ch}
\and
\IEEEauthorblockN{Dimitrios Dimitriadis}
\IEEEauthorblockA{\textit{Microsoft Research} \\
dimitrios.dimitriadis@microsoft.com}
\linebreakand 
\IEEEauthorblockN{Shruti Tople}
\IEEEauthorblockA{\textit{Microsoft Research} \\
shruti.tople@microsoft.com}
\and
\IEEEauthorblockN{Robert West}
\IEEEauthorblockA{\textit{EPFL} \\
robert.west@epfl.ch}
}

\maketitle

\begin{abstract}
A large body of work shows that machine learning (ML) models can leak sensitive or confidential information about their training data.
Recently, leakage due to distribution inference (or property inference) attacks is gaining attention. In this attack, the  goal of an adversary is to infer distributional information about the training data.
So far, research on distribution inference has focused on demonstrating successful attacks, with little attention given to identifying the potential causes of the leakage and to proposing mitigations.
To bridge this gap, as our main contribution, we theoretically and empirically analyze the sources of information leakage that allows an adversary to perpetrate distribution inference attacks.
We identify three sources of leakage: (1) memorizing specific information about the \(\mathbb{E}[Y|X]\) (expected label given the feature values) of interest to the adversary, (2) wrong inductive bias of the model, and (3) finiteness of the training data.
Next, based on our analysis, we propose principled mitigation techniques against distribution inference attacks.
Specifically, we demonstrate that causal learning techniques are more resilient to a particular type of distribution inference risk termed \emph{distributional membership inference} than associative learning methods.
And lastly, we present a formalization of distribution inference that allows for reasoning about more general adversaries than was previously possible.
\end{abstract}

\section{Introduction}

Machine learning (ML) has conquered many applications where data is available, including applications with sensitive or confidential data. This is problematic if the trained ML models are exposed to other parties, since it has been shown that models leak information about their training data\cite{shokri2017membership,ateniese2015hacking}. Exposure to other parties happens due to various reasons. For example, the models are deployed in user software such as for image processing or next\hyp word\hyp prediction on smartphones \cite{apple2022core,hard2018federated}, or they are shipped with medical devices \cite{benjamens2020state}. They are often shared online or made available via API access as well \cite{caffe2022model,openai2022api}. Exposing the model to other parties can further be a side effect of the training, such as in federated learning \cite{mcmahan2017communication}, where the model is trained by multiple data owners collaboratively, who need access to the model in order to provide updates based on their local data.

Recently, a rising privacy concern is the problem of distribution inference (also called property inference) attacks \cite{ateniese2015hacking, ganju2018property,zhang2021leakage,suri2021formalizing}. These attacks assume an adversary who has access to a model that was trained on confidential data and wants to use their access to the model to infer information about the training distribution. That is, the adversary is not interested in individual records or small sets of records in the training dataset, but in distributional properties such as the proportion of males or the mean age of individuals in the training distribution.
For example, an e\hyp commerce company might try to learn which of a competitor's products are the most popular from a recommender system that this competitor has built. Or an attacker might try to determine whether an organization's systems are vulnerable to Meltdown or Spectre attacks by analyzing a model that this organization has trained to detect cryptomining activity on their systems, as showcased in experiments by Ganju et al.\ \cite{ganju2018property}.

The problem of distribution inference was first identified  by Ateniese et al.\ \cite{ateniese2015hacking}. Since then various attacks for both the white\hyp box setting (the adversary has access to the model parameters) and the black\hyp box setting (the adversary only has query access to the model) have been proposed. These attacks have been shown to be successful against support vector machines and hidden Markov models \cite{ateniese2015hacking}, fully connected neural networks \cite{ganju2018property}, convolutional neural networks \cite{suri2021formalizing} and generative adversarial networks \cite{zhou2021property}.
However, defenses against distribution inference have received little attention, and the few proposed solutions have shortcomings that limit their use. For instance, some of them do not protect against the most common setting of black\hyp box adversaries. (See \Secref{sec:related-distribution_inference} for an overview.) We ascribe this partly to the lack of a systematic study of \emph{why} the attacks are successful, i.e., what makes models leak information about their training distribution. This is crucial for understanding in which settings such distribution leakage is unavoidable and in which settings it can be prevented, and ultimately building defenses against distribution inference attacks by closing the sources of leakage.

\xhdr{Our approach}
In this paper, we ask the question: \emph{Why do ML models leak information about their training data distribution?} We answer this question both theoretically and empirically. We first identify the different possible sources of leakage: (1) memorizing pieces of information about the function \(\mathbb{E}[Y|X]\) that the adversary is interested in, (2) inductive biases in the specification of the architecture or the training algorithm of the target model that do not agree with the training data, and (3) the finiteness of the training dataset. Our theoretical analysis (\Secref{sec:theory-reasons}) then relies on existence proofs by example: for each of the possible sources of leakage we show that there exists at least one ML model and one data distribution such that the model leaks information about its training distribution via this source. This implies that any defense mechanism that aims at offering protection in the worst\hyp case has to take all of these sources into account. We further perform state\hyp of\hyp the\hyp art distribution inference attacks on neural networks trained on synthetic data that is carefully crafted to isolate the different possible sources of leakage (\Secref{sec:experiments-distribution_inference})\footnote{The code for reproducing all experiments in this paper can be found at \url{https://github.com/epfl-dlab/distribution-inference-risks}.}. Our experiments show that information about the training distribution is leaked via all of these sources not only in theoretical examples, but also in more realistic settings, and that distribution inference attacks can pick up on the signal from all of these sources.
With our insights about the different sources of leakage, we are able to propose systematic defense strategies that are each aimed at closing one of the sources of leakage (\Secref{sec:defenses-distribution_inference}).

We further introduce an attack scenario termed \emph{distributional membership inference}, which combines distribution inference and membership inference (\Secref{sec:def-membership_inference}). It assumes an adversary that wants to know whether a particular party has contributed records to the training dataset, but instead of knowing the specific records of each party only knows the distributions that these records were drawn from. Based on our insights from the leakage analysis, we show that in some cases there is potential for protecting against such an adversary without a loss in utility when training the model via 
causal learning techniques such as invariant risk minimization (IRM) \cite{arjovsky2019invariant} as opposed to the traditional associational learning methods (\dblsecref{sec:defenses-membership_inference}{sec:experiments_causal}).

Along the way, we generalize the formalization of binary distribution inference by Suri and Evans \cite{suri2021formalizing} to adversaries with arbitrary hypotheses, which allows, e.g., for a formal treatment of adversaries that perform regression instead of binary classification attacks (\Secref{sec:definition-distribution_inference}). We further release a Python library that implements state\hyp of\hyp the\hyp art white\hyp box and black\hyp box distribution inference attacks\cite{meynet2022pia}. We hope that both of these contributions will serve as tools for future analyses of the distribution inference setting and for the development of new attacks and defenses.

\xhdr{Contributions} In summary, our contributions are these:
\begin{itemize}
    \item We theoretically and empirically identify the sources of distribution leakage. Based on our analysis we propose principled mitigation strategies against distribution inference attacks.
    \item We introduce \emph{distributional membership inference} and show how causal techniques can offer protection against it.
    \item We present a more general formalization of distribution inference.
    \item We release a Python library that implements state\hyp of\hyp the\hyp art distribution inference attacks.
\end{itemize}

\section{Background}
\label{sec:background}
We introduce some background that we will need later in the paper.

\subsection{Distribution inference attacks}
\label{sec:background-distribution_inference}
The problem of distribution inference has first been described by Ateniese et al.\ \cite{ateniese2015hacking} as a problem of binary inference: there exist two possible training distributions \(\mathcal{D}^0\) and \(\mathcal{D}^1\). The adversary knows that the target model was trained on one of them, and tries to figure out on which one. Typically, these two distributions differ in the proportion of a binary attribute that may or may not be one of the features used during training. The binary setting is the one considered in most of the prior work, with the exception of the work by Zhang et al.\ \cite{zhang2021leakage}, which also considers a finite set of possible training distributions, and the works by Zhou et al.\ \cite{zhou2021property} and Suri and Evans \cite{suri2021formalizing}, which perform regression attacks on continuous properties of the training distribution. To also cover these more general settings, we will index the training distribution \(\mathcal{D}^r\) by a variable \(r \in R\), which in the regression setting denotes the regression target and in the binary setting can take the value \(0\) or \(1\).

Many attacks rely on meta\hyp classifiers whose output is a guess for \(r\). In the white\hyp box setting, the input to these meta\hyp classifier are the weights of the target model. In the black\hyp box setting, the input are outputs of the target model. To generate training data for the meta\hyp classifiers, the adversary trains so\hyp called \emph{shadow models}. These are models with the same architecture and training parameters as the target model that are trained on samples from \(\mathcal{D}^r\) for different values of \(r\), and labeled by that value, akin to the shadow models used for membership inference attacks \cite{shokri2017membership}. A white\hyp box meta\hyp classifier can then be trained by using the weights of the shadow models as training features, and a black\hyp box meta\hyp classifier by using outputs of the shadow models.

\subsection{Invariant representation learning}
Classic associational ML models trained via \textit{empirical risk minimization} (ERM)  predict a label \(Y\) from a list of features \(X\) whilst trying to minimize the prediction error on their training set. This is achieved by taking all features from \(X\) that are predictive of \(Y\) into account. However, the relationship of some of these features with \(Y\) might be different in the train than in the test distribution. Take the example by Arjovsky et al.\ \cite{arjovsky2019invariant} of a model that is supposed to distinguish pictures of cows from pictures of camels. It might very well happen that in the training data the vast majority of cows is shown in front of a grassy background, whereas most camels are shown in a sandy desert. A classic ERM model might then base its predictions purely on the background of the picture and thereby achieve high accuracy on the training data. However, the correlation between the background and the type of animal is a spurious one: there exist camels that do not stand in a desert. This is why at test time the model might be confronted with pictures of cows in front of sandy backgrounds and camels in front of grassy backgrounds, for which its predictions will then be completely wrong.

Recently, works on the theory of causality \cite{pearl, abs-1911-10500} have argued that out-of-distribution generalization must consider the causal nature of the underlying data generation mechanism. 
Causal relationships are the ones expected to be robust and generalizable \cite{icc}, since they are not spurious artifacts of the data. In our previous example, the property of having a hump would be a causal feature of a camel. Unfortunately, for problems of interest the underlying causal model is often unknown. Causal ML is a field aiming to develop techniques that can still capture some causal properties of the data.
An important idea of causal ML is the \emph{invariance principle}, which states that only relationships stable across different training distributions should be preserved \cite{icc, inv_feature_rep}. Indeed, under certain assumptions the invariant relationships are expected to be the ones between the target variable and its causal parents \cite{arjovsky2019invariant}.
The different training distributions alluded to earlier correspond to data collected in different environments $e \in \mathcal{E}_{\tr}$, e.g., data contributed by people living in different geographical regions. Each environment $e \in \mathcal{E}_{\tr}$ induces i.i.d.\ samples from a distribution that is specific to this environment.

An implementation approach for the abstract invariance principle is proposed by Arjovsky et al.\ \cite{arjovsky2019invariant}. The authors introduce \emph{invariant risk minimization} (IRM), an alternative to ERM, and a practical training objective enforcing invariance in a learned latent representation.
IRM aims at finding a predictor $f(X) \approx Y$ that performs well across the set of  environments $\mathcal{E}^*$, only part of which were seen during training:
$\mathcal{E}_{\tr} \subsetneq \mathcal{E}^*$.
To achieve this, the authors decompose $f$ into a representation learning component $\Phi$ and a model $w$, such that $f = w \circ \Phi$, where $\circ$ denotes function composition.
The feature representation $\Phi$ is deemed invariant if the same model $w$ is simultaneously optimal for all environments $e \in \mathcal{E}_{\tr}$. 
Intuitively, $\Phi$ is an invariant representation if its representation is equally useful for all environments.

Thus, IRM solves the following optimization problem:
\begin{align*}
    &\min\limits_{\Phi, w} \sum\limits_{e \in \mathcal{E}_{\tr}} R^e(w \circ \Phi), \\
    &\text{subject to } w \in \arg\min\limits_{w'} R^e(w' \circ \Phi), \text{ for all } e \in \mathcal{E}_{\tr},
\end{align*}
where $R^e$ is the empirical risk computed within environment $e$. In this paper we show that IRM can leak less distributional information about its training data than ERM.

\section{Definitions}

We begin by defining the problems of distribution inference and distributional membership inference, with the most frequent symbols used in this paper explained in Table~\ref{fig:glossary}.

\begin{tablefig}
\centering
\begin{tabular}{ |p{0.13\linewidth}|p{0.73\linewidth}| } 
\hline
\textbf{Symbol} & \textbf{Meaning}\\
\hline
\(\Phi\) & Representation learner of IRM \\
\(w\) & Prediction model on top of \(\Phi\) \\
\(R\) & Set of distribution indices \\ 
\(r\) & Index of the training distribution \\ 
\(\mathcal{D}^r\) & Training distribution \\
\(X^r, Y^r\) & Feature and label random variables distributed according to \(\mathcal{D}^r\) \\
\(M\) & Target model \\ 
\(\hat{r}\) & Guess for \(r\) by the adversary \\
\(d\) & Function on \(R\times R\) to asses the quality of the guess \(\hat{r}\) \\
\(\mathcal{D}^i_{\party}\) & Distribution of party \(i\)'s data in the membership inference setting \\
\(\beta\) & Vector of linear regression parameters \\
\(\hat{\beta}\) & Estimator for \(\beta\) \\
\hline
\end{tabular}
\caption{Glossary for the symbols used in this paper.}
\label{fig:glossary}
\end{tablefig}

\subsection{Distribution inference}
\label{sec:definition-distribution_inference}
\begin{figure}
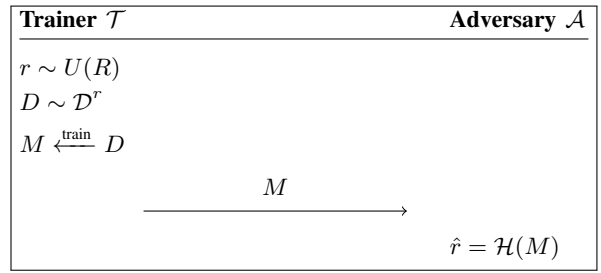

\centering
\begin{pcvstack}[center,boxed]
\pseudocodeblock{
\textbf{Trainer \(\mathcal{T}\)} \> \> \textbf{Adversary \(\mathcal{A}\)} \\
[0.1\baselineskip][\hline] \<\<\\[-0.5\baselineskip]
r\sim U(R)  \> \>\\
D\sim\mathcal{D}^r \> \>\\
M\xleftarrow{\text{train}} D \> \>\\
\> \sendmessageright*{M} \> \\
\> \> \hat{r}=\mathcal{H}(M) }
\end{pcvstack}
\caption{\textbf{Distribution inference game.} The training distributions are indexed by a value \(r\). The trainer \(\mathcal{T}\) first samples a distribution index \(r\) and then samples a training set \(D\) from \(\mathcal{D}^r\), on which they train a model \(M\). By applying an algorithm \(\mathcal{H}\) to \(M\), the adversary computes a guess \(\hat{r}=\mathcal{H}(M)\) for \(r\).}
\label{fig:distribution_inference_game}
\end{figure}

We define \textit{distribution inference} as a cryptographic game between the adversary \(\mathcal{A}\) and the model trainer \(\mathcal{T}\) (\Figref{fig:distribution_inference_game}) as a generalization of the binary definition by Suri and Evans \cite{suri2021formalizing}; below we describe how their definition relates to ours. We allow for an arbitrary, even uncountable number of potential training distributions \(\mathcal{D}^r\) indexed by a set \(R\). The trainer samples an index \(r\) from \(R\) uniformly at random
and then samples a training set \(D\) from \(\mathcal{D}^r\). Finally, the model \(M\) is trained on \(D\). The adversary applies an algorithm \(\mathcal{H}\) to the model \(M\) that returns a guess \(\hat{r}\) about the training distribution. We explicitly write \(\hat{r}|r\) for the guess when the true index is \(r\) to avoid ambiguities in the following formulas.

Let \(d\) be a (not necessarily symmetric) function that measures distances between objects in \(R\). The goal of the adversary is to minimize \(d(r,\hat{r}|r)\), the distance between the true index \(r\) and the guess \(\hat{r}|r\). A natural measure for the performance of an attack algorithm is thus the expected distance
\begin{equation*}
    \mathbb{E}[d(r,\hat{r}|r)].
\end{equation*}
The expectation is taken over the random sampling of \(r\), over the random sampling of \(D\), and over the potential randomness in the training of \(M\) and in \(\mathcal{H}\). 
We define the \emph{advantage of the adversary} \(\mathcal{A}\) using algorithm \(\mathcal{H}\) as the expected improvement over the optimal guess without access to \(M\):
\begin{equation*}
    \adv_{\mathcal{H}} = \mathbb{E}[d_0-d(r,\hat{r}|r)],
\end{equation*}
where
\begin{equation*}
    d_0 = \inf_{r'\in R}\mathbb{E}[d(r,r')].
\end{equation*}

We can recover Suri and Evans' definition of the cryptographic game and of the adversarial advantage (up to a sign) by setting \(R=\{0,1\}\) and \(d(r, r') = \abs{r-r'}\).
In the rest of this paper, whenever we work in a binary setting with two distributions, we mean this choice of \(d\). We further use the notation \((X^b,Y^b)\sim \mathcal{D}^b\) for the random variables corresponding to the features and the label distributed according to distribution \(\mathcal{D}^b\), for \(b=0,1\).

However, our definition goes beyond the binary case. For instance, it also covers adversaries that perform regression to infer a continuous property of the training distribution. Consider, e.g., the task of inferring the ratio of females among the customers of the e\hyp commerce company that builds a recommender system in the example from the introduction. This ratio lies in the interval \([0,1]\) and thus we set \(R=[0,1]\). For \(d\) we can, e.g., again choose the absolute difference \(d(r, r') = \abs{r-r'}\) (one alternative would be the squared distance). For these choices, we have $d_0 = \frac{1}{4}$, where the infimum is attained at \(r'=\frac{1}{2}\).

Note that our definition also allows for arbitrary sets of distributions \(\{\mathcal{D}^r\}_{r\in R}\). We can define \(d\) as
\begin{equation*}
    d(r,r') = \tilde{d}(\mathcal{D}^r,\mathcal{D}^{r'}),
\end{equation*}
where \(\tilde{d}\) is a divergence or metric on the space of distributions, such as the KL divergence or the Wasserstein distance.

\subsection{Distributional membership inference}
\label{sec:def-membership_inference}
Many training datasets in ML consist of data that was generated by multiple parties: each transaction in a financial dataset was initiated by one legal entity, each post in a social network was created by one individual, each bike ride in a location dataset was done by one individual. 
We can view the data points contributed by a given party as samples from a distribution specific to this party. 
We consider an adversary that we consider wants to learn whether one particular party contributed to the training dataset or not.

Formally, we assume that there are \(n\) parties. For some fixed index \(i_0\), the adversary wants to know whether party \(i_0\) contributed to the training dataset. To each party \(i\) corresponds one distribution \(\mathcal{D}^i_{\party}\), and one dataset \(D^i_{\party}\) that is sampled from \(\mathcal{D}^i_{\party}\). 
We define the following two datasets:
\begin{align*}
    D^0 = &\bigcup_{i=1}^n D^i_{\party}, \\
    D^1 = &\bigcup_{i=1, i\neq i_0}^n D^i_{\party}.
\end{align*}
Thus, $D^0$ and $D^1$ are such that they differ only in the contribution of party $i_0$: one dataset contains party $i_0$'s data and the other does not. Implicitly, $D^0$ is sampled according to a distribution $\mathcal{D}^0$ and $D^1$ is sampled according to a distribution $\mathcal{D}^1$.
The model trainer \(\mathcal{T}\) and the adversary \(\mathcal{A}\) now play the distribution inference game in \Figref{fig:distribution_inference_game} with the two distributions \(\mathcal{D}^0\) and \(\mathcal{D}^1\): \(\mathcal{T}\) trains a model \(M\) either on data sampled from \(\mathcal{D}^0\) or on data sampled from \(\mathcal{D}^1\), and \(\mathcal{A}\) tries to determine on data from which of the two distributions \(M\) was trained. We call this problem \emph{distributional membership inference}. If there is no party \(i,\ i\neq i_0,\) with the same distribution as party \(i_0\) (\ie, \(\mathcal{D}^i_{\party} \neq \mathcal{D}^{i_0}_{\party}\) for all \(i\neq i_0\)), then this problem equivalent to determining whether party \(i_0\) contributed to the training dataset or not.

\xhdr{Comparison with record\hyp based membership inference}
Typically, membership inference in ML is investigated on a record level \cite{shokri2017membership}. That is, each party is assumed to have contributed at most one record to the training dataset. By observing the model, the adversary tries to learn whether a particular party, whose record they know, has contributed to the dataset or not.
In distributional membership inference each party might have contributed multiple records. Even though it is not the standard setting in membership inference, some works such as those on DP can also be extended to allow for multiple records per party \cite{dwork2008theory} (see also \Secref{sec:related}). The other difference between distributional membership inference and most prior work on membership inference is that distributional membership inference assumes a weaker adversary that does not know the exact records contributed by the target party, but only the distribution that they were sampled from. One example for such a setting is the following:
Consider an adversary that wants to know whether a particular person has contributed to a database of cancer patients, and thus has cancer. The records contributed by each person consist of measurements of this person's weight, blood pressure (and potentially other health markers) over multiple days. The adversary might have information about the target person's weight and blood pressure. Since weight and blood pressure fluctuate from day to day within the same individual, it is reasonable to assume that the adversary does not know the person's exact weight and blood pressure on the different days, but only the distribution of these values. The distribution of the weight could, e.g., be estimated by an estimated mean based on the visual appearance of the person plus Gaussian noise, and the distribution of the blood pressure by an estimated mean based on the (estimated) weight and the person's fitness level plus Gaussian noise.

\xhdr{Relation to the work by Suri et al.\ \cite{suri2022subject}}
In concurrent and independent work, Suri et al.\ \cite{suri2022subject} propose the notion of subject membership inference. They consider the same attack objective as us, but study it in the context of subject\hyp level membership inference in a cross\hyp silo federated learning (FL) setting. The term 'subject' usually refers to individual people. The authors assume an adversary that wants to determine whether a particular subject's data was used in the federated training of an ML model, where each subject can have contributed to the datasets of multiple parties that participate in the training. We note that our proposed defense against distributional membership inference based on IRM (\Secref{sec:defenses-membership_inference}) also extends to the FL setting.

\section{Reasons for distribution leakage}
\label{sec:theory-reasons}
What are the reasons why models leak information about their training distribution? We answer this question theoretically and show how all of the different potential sources of leakage can in fact leak distributional information. We show that this is already the case for such a simple model as linear regression, and thus all of these sources already have to be taken into account even in simple cases. In \Secref{sec:experiments-distribution_inference} we additionally show, for the case of neural networks, that not only theoretical attacks, but also the practical attacks proposed in prior work can exploit all of these sources of leakage.

For simplicity we will assume throughout this section the binary setting where \(R=\{0,1\}\), i.e., an adversary that wants to know whether the training distribution was either \(\mathcal{D}^0\) or \(\mathcal{D}^1\), for two distributions \(\mathcal{D}^0\), \(\mathcal{D}^1\). Our results extend to the regression case; see the end of this subsection.

Since we base our examples on linear regression target models, we give some background on linear regression. The assumption underlying linear regression is that the relationship between the features \(X=(X_1,\dots,X_d)\) and the label \(Y\) takes the form
\begin{equation*}
    Y = \beta_0 + \beta_1 X_1 + \dots + \beta_d X_d + \varepsilon
\end{equation*}
for a fixed vector \(\beta\) and a Gaussian error term \(\varepsilon\sim N(0,\sigma^2)\).
For an \(n\times (d+1)\)\hyp matrix \(\tilde{X}\) whose rows are independent samples of \((1, X_1, \dots, X_d)\), a vector of independent error terms \(\varepsilon_i\), \(i=1,\dots,n\), and the corresponding vector of labels \(\tilde{Y}\), a least\hyp squares linear regression model consists of the minimizer \(\hat{\beta}\) of the loss
\begin{equation*}
    \norm{\tilde{X}\hat{\beta} - \tilde{Y}}^2,
\end{equation*}
given by
\begin{equation*}
    \hat{\beta} = (\tilde{X}^T \tilde{X})^{-1} \tilde{X}^T \tilde{Y}.
\end{equation*}
Under the aforementioned assumption on the data distribution, \(\hat{\beta}\) converges to \(\beta\) when increasing the number of samples, which implies that
\begin{equation*}
    \mathbb{E}[\hat{\beta}] = \beta.
\end{equation*}
Further,
\begin{equation*}
    \var[\hat{\beta}] = \mathbb{E}_{\tilde{X}}\left[\sigma^2 (\tilde{X}^T\tilde{X})^{-1}\right].
\end{equation*}
In the one\hyp dimensional case \(d=1\) we get for the variance of \(\hat{\beta}_1\):
\begin{equation*}
    \var[\hat{\beta}_1] = \mathbb{E}_{\tilde{X}}\left[\frac{\sigma^2}{n s(\tilde{X})^2}\right],
\end{equation*}
where
\begin{equation*}
    s(\tilde{X})^2 = \frac{1}{n} \sum_{i=1}^n \left(\tilde{X}_{i1} - \frac{1}{n} \sum_{j=1}^n \tilde{X}_{j1}\right)^2
\end{equation*}
is the sample variance of the features.
Note that in these variances the expectation w.r.t.\ the error terms \(\varepsilon_i\), which are by assumption normally distributed, has been computed --- resulting in the factor \(\sigma^2\) --- whereas the expectation w.r.t.\ the features \(\tilde{X}\) depends on the feature distribution.

\xhdr{Proof technique}
The vector \(\hat{\beta}\) is a random variable, where the randomness comes from the sampling of the training data \((\tilde{X},\tilde{Y})\). In the following analyses we will show for different pairs \(\mathcal{D}^0\), \(\mathcal{D}^1\) that the distribution of \(\hat{\beta}\) differs when sampling the training data from \(\mathcal{D}^0\) vs.\ sampling it from \(\mathcal{D}^1\). Denote by \(\mathcal{D}_{\hat{\beta}}^0\) the distribution of \(\hat{\beta}\) when sampling the training data from \(\mathcal{D}^0\), and by \(\hat{\beta}^0\) the corresponding random variable. Define \(\mathcal{D}_{\hat{\beta}}^1\) and \(\hat{\beta}^1\) analogously. In the different settings that we analyze we will either show that the means of \(\mathcal{D}_{\hat{\beta}}^0\) and \(\mathcal{D}_{\hat{\beta}}^1\) differ or that their variances differ. This implies that there is a non\hyp null set \(V\) of \(\hat{\beta}\) values that have a different likelihood under \(\mathcal{D}_{\hat{\beta}}^0\) than under \(\mathcal{D}_{\hat{\beta}}^1\).
The adversary in \Figref{fig:distribution_inference_game} receives a sample \(\hat{\beta}^b\). If this sample comes from \(V\), they can predict with an accuracy that is strictly better than random guessing whether the sample came from \(\mathcal{D}_{\hat{\beta}}^0\) or \(\mathcal{D}_{\hat{\beta}}^1\). Thus, the advantage of the optimal adversary is strictly greater than \(0\) and hence the model leaks distributional information.

For simplicity we focus on the binary classification setting. However, our results can easily be extended to the regression setting where for instance \(R=[0,1]\) and the first regression parameter of \(\mathcal{D}^r\) is \(\beta^r_1=r\) in the example in \Secref{sec:theory-reasons-y_given_x}, or the parameter \(c\) in the examples in \Secref{sec:theory-reasons-imperfect_models} is \(c=1+r\) for distribution \(\mathcal{D}^r,\ r>0\).
\subsection{Reason 1: Differences in \(\mathbb{E}[Y|X]\)}
\label{sec:theory-reasons-y_given_x}
The setting in which it is maybe not surprising that the target model can leak information about the training distribution is the one in which \(\mathbb{E}[Y^0|X^0]\neq \mathbb{E}[Y^1|X^1]\), i.e., where there exists a feature vector \(x\) such that
\begin{equation*}
    \mathbb{E}[Y^0|X^0=x]\neq \mathbb{E}[Y^1|X^1=x].
\end{equation*}
Consider the e\hyp commerce company from the beginning, and assume that it sells sports products, including surfboards. As part of its recommender system it builds a model to predict how likely a particular client is to buy a particular product, where one of the features is the time of year. A competing company might want to determine which region of the world their competitor's customers are located in. If we expect that people are more likely to buy surfboards in summer than in winter, then for customers in the northern hemisphere the months June, July and August would have a positive correlation with the label, whereas for customers in the southern hemisphere they would have a negative correlation with the label.

Assume that \(\mathcal{D}^0\) and \(\mathcal{D}^1\) fulfill the linear regression assumptions with vectors \(\beta^0\) and \(\beta^1\). Since \(\mathbb{E}[Y^0|X^0]\neq \mathbb{E}[Y^1|X^1]\) and the error terms are assumed to have zero mean, we have that \(\beta^0 \neq \beta^1\). Hence,
\begin{equation*}
    \mathbb{E}[\hat{\beta}^0] = \beta^0 \neq \beta^1 = \mathbb{E}[\hat{\beta}^1].
\end{equation*}
\subsection{Reasons 2 \& 3: Imperfect models}
\label{sec:theory-reasons-imperfect_models}
In this subsection we assume that \(\mathbb{E}[Y^0|X^0]=\mathbb{E}[Y^1|X^1]\). In order for the distributions \(\mathcal{D}^0\) and \(\mathcal{D}^1\) to differ, the marginal distributions of the features \(\Pr(X^0)\), \(\Pr(X^1)\) and/or the marginal distributions of the label \(\Pr(Y^0)\), \(\Pr(Y^1)\) need to differ. In the following examples both the distribution of the features and the distribution of the labels differs between \(\mathcal{D}^0\) and \(\mathcal{D}^1\). Consider again the example of the e\hyp commerce company that sells sports products. A competitor might be interested in knowing whether that company has many or few young customers. If the age of the user is one of the features that the model uses and we assume that surfboards are more popular among younger than among older users, then in this example both the distribution of the features and the distribution of the label would differ between \(\mathcal{D}^0\) and \(\mathcal{D}^1\).

If \(\mathbb{E}[Y^0|X^0]=\mathbb{E}[Y^1|X^1]\), a model that perfectly learns the relationship between features and label --- i.e., a model that learns exactly the function \(g(x)=\mathbb{E}[Y^0|X^0=x]\) --- would not leak any information about whether it was trained on \(\mathcal{D}^0\) or \(\mathcal{D}^1\). However, models in the real world are not perfect. There are two reasons for this: the fact that there are only finitely many training samples and a wrong inductive bias, i.e., assumptions made in the choice of model architecture or training algorithm that are not fulfilled by the data and lead to a model that does not learn the correct function \(\mathbb{E}[Y|X]\). As we show in this section, both of these sources of imperfection can cause a model to leak information about its training distribution, even if \(\mathbb{E}[Y^0|X^0]=\mathbb{E}[Y^1|X^1]\).

\subsubsection{Reason 2: Wrong inductive bias}
\label{sec:theory-inductive_biases}
Consider distributions \(\mathcal{D}^0\), \(\mathcal{D}^1\) with one\hyp dimensional feature vectors and the labels
\begin{align*}
    Y^0 &= X_1^0 + 1 + \varepsilon\\
    Y^1 &= X_1^1 + 1 + \varepsilon.
\end{align*}
Assume that \(\mathcal{D}^0\) and \(\mathcal{D}^1\) only differ in the marginal distribution of \(X_1\) (and therefore also of \(Y\)): assume that \(X_1^0\) has a probability density \(f^0\), and define the density \(f^1\) of \(X^1\) as \(f^1(cx) = f^0(x)\) for a factor \(c>1\); informally, \(X_1^1\sim c X_1^0\).
We assume that the target model is a linear regression model without an intercept term, i.e.,
\begin{equation*}
    Y = \beta_1 X_1 + \varepsilon.
\end{equation*}
It thus induces a wrong bias
and cannot correctly model the data distribution. The minimizer \(\hat{\beta}\) for the loss of a one\hyp dimensional linear regression model without an intercept term is given by
\begin{equation*}
    \hat{\beta}_1 = \frac{\sum_i \tilde{X}_{i1} \tilde{Y}_i}{\sum_i \tilde{X}_{i1}^2}.
\end{equation*}
We have
\begingroup
\allowdisplaybreaks 
\begin{align*}
    &\mathbb{E}[\hat{\beta}_1^1]\\
    &\overset{\hphantom{\mathbb{E}[\varepsilon_i]=0}}{=} \mathbb{E}_{\{\tilde{X}_{i1},\tilde{Y}_i\}_{i=1}^n\sim\mathcal{D}^1}\mathbb{E}_{\{\varepsilon_i\}_{i=1}^n\sim N(0,\sigma^2)}(\hat{\beta}_1)\\
    &\overset{\hphantom{\mathbb{E}[\varepsilon_i]=0}}{=} \mathbb{E}_{\{\tilde{X}_{i1},\tilde{Y}_i\}_{i=1}^n\sim\mathcal{D}^1}\mathbb{E}_{\{\varepsilon_i\}_{i=1}^n}\left(\frac{\sum_i \tilde{X}_{i1} \tilde{Y}_i}{\sum_i (\tilde{X}_{i1})^2}\right)\\
    &\overset{\hphantom{\mathbb{E}[\varepsilon_i]=0}}{=} \mathbb{E}_{\{\tilde{X}_{i1},\tilde{Y}_i\}_{i=1}^n\sim\mathcal{D}^1}\mathbb{E}_{\{\varepsilon_i\}_{i=1}^n}\left(\frac{\sum_i \tilde{X}_{i1} (\tilde{X}_{i1} + 1+\varepsilon_i)}{\sum_i (\tilde{X}_{i1})^2}\right)\\
    &\overset{\mathbb{E}[\varepsilon_i]=0}{=} 1 + \mathbb{E}_{\{\tilde{X}_{i1},\tilde{Y}_i\}_{i=1}^n\sim\mathcal{D}^1}\left(\frac{\sum_i \tilde{X}_{i1} }{\sum_i (\tilde{X}_{i1})^2}\right) + 0\\
    &\overset{\hphantom{\mathbb{E}[\varepsilon_i]=0}}{=} 1 + \mathbb{E}_{\{\tilde{X}_{i1},\tilde{Y}_i\}_{i=1}^n\sim\mathcal{D}^0}\left(\frac{\sum_i c\tilde{X}_{i1} }{\sum_i (c\tilde{X}_{i1})^2}\right)\\
    &\overset{\hphantom{\mathbb{E}[\varepsilon_i]=0}}{=} 1 + c\mathbb{E}_{\{\tilde{X}_{i1},\tilde{Y}_i\}_{i=1}^n\sim\mathcal{D}^0}\left(\frac{\sum_i \tilde{X}_{i1} }{\sum_i (\tilde{X}_{i1})^2}\right)\\
    &\overset{\hphantom{\mathbb{E}[\varepsilon_i]=0}}{=} \mathbb{E}_{\{\tilde{X}_{i1},\tilde{Y}_i\}_{i=1}^n\sim\mathcal{D}^0}\mathbb{E}_{\{\varepsilon_i\}_{i=1}^n}(\hat{\beta}_1)\\
    &\hphantom{\overset{\mathbb{E}[\varepsilon_i]=0}{=}}+ (c-1) \mathbb{E}_{\{\tilde{X}_{i1},\tilde{Y}_i\}_{i=1}^n\sim\mathcal{D}^0}\left(\frac{\sum_i \tilde{X}_{i1} }{\sum_i (\tilde{X}_{i1})^2}\right)\\
    &\overset{\hphantom{\mathbb{E}[\varepsilon_i]=0}}{=} \mathbb{E}[\hat{\beta}_1^0] + (c-1) \mathbb{E}_{\{\tilde{X}_{i1},\tilde{Y}_i\}_{i=1}^n\sim\mathcal{D}^0}\left(\frac{\sum_i \tilde{X}_{i1} }{\sum_i (\tilde{X}_{i1})^2}\right).
\end{align*}
\endgroup
Hence, unless \(X_1^0 \equiv 0\) (and therefore also \(X_1^1 \equiv 0\)), the expected value of the model parameter \(\hat{\beta}_1\) under \(\mathcal{D}^0\) and \(\mathcal{D}^1\) differs. The parameter of a model trained on \(\mathcal{D}^0\) converges to a different value than the parameter of a model trained on \(\mathcal{D}^1\) when increasing the size of the training dataset, and therefore leaks information about the training distribution.

\subsubsection{Reason 3: Finite amount of training data}
\label{sec:theory-finite_data}
Assume that a training set of size \(n\) is sampled either from \(\mathcal{D}^0\) or from \(\mathcal{D}^1\). Assume that both \(\mathcal{D}^0\) and \(\mathcal{D}^1\) are one\hyp dimensional linear distributions with the same \(\beta\) coefficients:
\begin{align*}
    Y^0 &= \beta_0 + \beta_1 X_1^0 + \varepsilon\\
    Y^1 &= \beta_0 + \beta_1 X_1^1 + \varepsilon,
\end{align*}
where, using the notation from the previous subsection, \(f^1(cx)=f^0(x)\) for some \(c>1\).
As we have seen earlier, the variance of the parameter \(\hat{\beta}_1\) is
\begin{equation*}
    \var[\hat{\beta}_1] = \mathbb{E}_{\tilde{X}}\left[\frac{\sigma^2}{n s(\tilde{X})^2}\right],
\end{equation*}
where \(s(\tilde{X})^2\) is the sample variance of the feature. We write \(\tilde{X}\sim f^b\) to denote a data matrix whose rows are sampled according to the density \(f^b\), for \(b=0,1\). Since \(f^1(cx) = f^0(x)\), we have that
\begin{align*}
    \var[\hat{\beta}_1^1] &= \mathbb{E}_{\tilde{X}\sim f^1}\left(\frac{\sigma^2}{n s(\tilde{X})^2}\right)\\
    &= \mathbb{E}_{\tilde{X}\sim f^0}\left(\frac{\sigma^2}{n s(c\tilde{X})^2}\right)\\
    &= \frac{1}{c^2} \mathbb{E}_{\tilde{X}\sim f^0}\left(\frac{\sigma^2}{n s(\tilde{X})^2}\right)\\
    &= \frac{1}{c^2} \var[\hat{\beta}_1^0].
\end{align*}
The lower variance of \(\hat{\beta}_1^1\) means that more extreme values of \(\hat{\beta}_1\) are more likely under \(\mathcal{D}^0\) than under \(\mathcal{D}^1\). Hence the model parameters leak information about the training distribution. However, when increasing the amount of training data \(n\), the variance of \(\hat{\beta}_1^0\) and \(\hat{\beta}_1^1\) will go to \(0\) both for \(\mathcal{D}^0\) and for \(\mathcal{D}^1\).

With the case \(\mathbb{E}[Y^0|X^0]\neq \mathbb{E}[Y^1|X^1]\), the problem of finitely many samples and the problem of a wrong inductive bias we have now covered all potential sources of leakage and shown they are actual sources or leakage.

\section{Defenses}
\label{sec:defenses}
In this section we use the insights from \Secref{sec:theory-reasons} to identify ways in which the leakage of information about the training distribution may be prevented, but also identify settings in which such a mitigation is impossible without sacrificing model performance. We do this both for classic distribution inference (\Secref{sec:defenses-distribution_inference}) and for distributional membership inference (\Secref{sec:defenses-membership_inference}).
\subsection{Mitigating distribution inference risks}
\label{sec:defenses-distribution_inference}
\subsubsection{Reducing the number of cases where \(\mathbb{E}[Y^0|X^0]\neq\mathbb{E}[Y^1|X^1]\)}
As we have seen in \Secref{sec:theory-reasons-y_given_x}, differences between \(\mathbb{E}[Y^0|X^0]\) and \(\mathbb{E}[Y^1|X^1]\) can lead to distribution leakage. This might arguably be the worst source of leakage, since in a setting where \(\mathbb{E}[Y^0|X^0]\neq\mathbb{E}[Y^1|X^1]\), a model that perfectly learns \(\mathbb{E}[Y^b|X^b]\), i.e., a perfect model from a pure ML perspective, exhibits the highest amount of leakage. This means that the prevention of leakage and the goal of training a model that fits the data well, e.g., by improving the model's inductive bias or increasing the amount of training data, are directly opposed --- a no\hyp free\hyp lunch situation. It gets even worse when considering that both improving the model's inductive bias and increasing the amount of training data in themselves are in isolation ways to \emph{decrease} the amount of leakage (see \dblsecref{sec:defenses-distribution_inference-inductive_biases}{sec:defenses-distribution_inference-training_data}).

It is therefore worth to first think about whether an adversary of interest, i.e., one  trying to infer information that the model trainer considers as confidential, exists for which \(\mathbb{E}[Y^0|X^0]\neq\mathbb{E}[Y^1|X^1]\). If this is not the case, then the model might still leak distributional information via this mechanism, but just information that is not considered confidential.

If the model trainer cannot rule out adversaries of interest for which \(\mathbb{E}[Y^0|X^0]\neq\mathbb{E}[Y^1|X^1]\), then they may defend against them, or at least against some of them, by reducing the number of features that the model relies on for its predictions. If, for instance, the model relies only on the first \(k\) of \(d\) features, for \(k<d\), then an adversary with \(\mathbb{E}[Y^0|X^0]\neq\mathbb{E}[Y^1|X^1]\) but \(\mathbb{E}[Y^0|X_1^0,\dots,X_k^0]=\mathbb{E}[Y^1|X_1^1,\dots,X_k^1]\) does not pose a problem anymore. Similar statements hold for models that do take all features into account, but reduce the dimensionality in a latent space, such as when performing preprocessing via principle component analysis. Even though, given enough training data, taking all features into account should give the best performing model, such dimensionality reduction methods often improve the model fit, e.g., by reducing overfitting. And even methods that do reduce the model fit on the available test data may be desirable for out\hyp of\hyp distribution generalization. Causal learning methods as discussed in \Secref{sec:related-causal_learning} fall into this category: they aim to learn models that only rely on those features that have a causal effect on the label \(Y\), and ignore features that are only spuriously correlated with the label. Let \(X_C\) denote the causal parents of \(Y\). Then one may even argue that no adversary of interest exists for which \(\mathbb{E}[Y^0|X_C^0]\neq\mathbb{E}[Y^1|X_C^1]\), since causal relationships can be seen as laws of the universe \cite{peters2017elements}, and thus, if \(\mathbb{E}[Y^0|X_C^0]\neq\mathbb{E}[Y^1|X_C^1]\), either \(\mathcal{D}^0\) or \(\mathcal{D}^1\) is not a realistic distribution. A causal learning method that learns the function \(\mathbb{E}[Y^b|X_C^b]\) would thus offer protection against the source of leakage discussed in this paragraph.

\subsubsection{Correcting a wrong inductive bias}
\label{sec:defenses-distribution_inference-inductive_biases}
The example in \Secref{sec:theory-inductive_biases} shows that a wrong inductive bias --- there in the form of a wrong model architecture --- can be a source of distribution leakage. This wrong bias may be corrected by thinking more carefully about the explicit and implicit assumptions that are made by the choice of model architecture and training algorithm (including data preprocessing); by testing some of these assumptions on the available training data; and by consulting with domain experts. All of these measures will then hopefully lead to a new set of assumptions that better match the data distribution and therefore reduce the distribution leakage of the resulting model. We note that obtaining an inductive bias that closely matches the data distribution is also a goal in a classic ML setting where one simply wants to improve the model fit. Thus, here the goals of model performance and of training distribution protection are aligned.

\subsubsection{Increasing the amount of training data}
\label{sec:defenses-distribution_inference-training_data}
In \Secref{sec:theory-finite_data} we show that the finiteness of the training dataset can cause the model trained on this dataset to leak information about the training distribution. A natural way to reduce this leakage is to collect more training data or to use data augmentation \cite{shorten2019survey,feng2021survey}. However, this should not be done as an isolated measure. Increasing the amount of training data will usually improve the model fit. But in the case where \(\mathbb{E}[Y^0|X^0]\neq\mathbb{E}[Y^1|X^1]\), a better fitted model will leak \emph{more} information about the training distribution rather than less. Similarly, in the case of a wrong inductive bias a better fitted model can leak more information than a worse fitted one; see our example in \Secref{sec:theory-finite_data}. It is thus important to first address these other sources of leakage before collecting more training data.
\subsection{Mitigating distributional membership inference risks via causal learning}
\label{sec:defenses-membership_inference}
As described in \Secref{sec:def-membership_inference}, in the distributional membership inference setting we assume that there are \(n\) parties and that the adversary knows that \(n-1\) of them --- w.l.o.g. parties \(1,\dots,n-1\) --- have contributed data to the training dataset. The adversary wants to determine whether the \(n\)\hyp th party has contributed data as well or not.

For the discussion in this subsection we only focus on the problem of reducing the leakage from \(\mathbb{E}[Y^0|X^0]\neq\mathbb{E}[Y^1|X^1]\), since this is the most challenging problem and the most interesting one in this setting, and since the problems of a wrong inductive bias and of the finiteness of the training data can be tackled using more standard ML techniques (see \Secref{sec:defenses-distribution_inference}).

In order for \(\mathbb{E}[Y^0|X^0]\neq\mathbb{E}[Y^1|X^1]\) to hold, it is necessary that \(\mathcal{D}^n_{\party}\) differs from \(\mathcal{D}^i_{\party}\) for at least one \(i\).
If the target model should be useful and achieve a low error on each of the distributions \(\mathcal{D}^i_{\party}\) --- we will call them training environments --- or even generalize beyond the training environments, one has to make certain assumptions on how the environments \(\mathcal{D}^i_{\party}\) may differ; arbitrarily differing training environments could lead to a model that performs arbitrarily badly on at least one of the training environments. One such assumption is made by IRM (see \Secref{sec:related-causal_learning}). This assumption is made to achieve out\hyp of\hyp distribution generalization of the trained model. In this section we are not interested in out\hyp of\hyp distribution generalization, but we rather show that this assumption at the same time also helps protect against distributional membership inference. The assumption is that for all \(i\), \(j\) it holds that
\begin{equation}
\label{eq:irm_assumption}
    \mathbb{E}_{(X,Y)\sim \mathcal{D}^i_{\party}}(Y|\Phi(X))=\mathbb{E}_{(X,Y)\sim \mathcal{D}^j_{\party}}(Y|\Phi(X)),
\end{equation}
where \(\Phi\) denotes an encoder that maps the set of all features to the causal parents of \(Y\). In the simplest case, these causal parents are a subset \(X_C\) of the features \(X\), and \(\Phi\) is a binary matrix that selects the features \(X_C\).
If the IRM training works as intended, then it learns the encoder \(\Phi\) and on top of it the optimal model \(w\) that outputs a prediction \(\hat{Y}\) of the label from the encoded features \(\Phi(X)\), i.e.,
\begin{equation*}
    \hat{Y} = (w \circ \Phi)(X).
\end{equation*}
Because of Assumption \ref{eq:irm_assumption} the optimality of \(w\) is invariant w.r.t.\ the environment, i.e., it is the optimal predictor on top of \(\Phi\) in every environment \(\mathcal{D}^i_{\party}\), since it only depends on the distribution \(\mathbb{E}[Y^b|\Phi(X^b)]\) and not on the distribution \(\mathbb{E}[Y^b|X^b]\). Assumption \ref{eq:irm_assumption} also implies that
\begin{equation*}
    \mathbb{E}[Y^0|\Phi(X^0)] = \mathbb{E}[Y^1|\Phi(X^1)].
\end{equation*}
Hence, \(w\) does not leak anything through the mechanism \(\mathbb{E}[Y^0|X^0]\neq\mathbb{E}[Y^1|X^1]\) --- at least in the black\hyp box case, since it might still be that adding or removing one environment changes the way that \(w\) computes the prediction function internally. If IRM works as intended and \(\Phi\) indeed computes the causal parents of \(Y\) (see Arjovsky et al.\ \cite{arjovsky2019invariant} for sufficient conditions in the linear case), then the composition \(w \circ \Phi\) also leaks nothing through this mechanism in the black\hyp box case. In our experiments in \Secref{sec:experiments_causal} we show that IRM can even provide a defense in the white\hyp box case.
This is as opposed to a standard ERM model that learns a model on top of the original features \(X\): Since we assume that \(\mathbb{E}[Y^0|X^0]\neq\mathbb{E}[Y^1|X^1]\), the optimal model --- the one that computes the function \(\mathbb{E}[Y^b|X^b]\) --- would be different for \(\mathcal{D}^0\) and \(\mathcal{D}^1\).

\section{Experiments}
\label{sec:experiments}
In this section we investigate experimentally whether the different sources of leakage that we identified in \Secref{sec:theory-reasons} are exploited by state\hyp of\hyp the\hyp art distribution inference attacks. We also test whether the defenses proposed in \Secref{sec:defenses} have the desired effect of reducing the performance of attacks.
In \Appref{app:subsampling} we analyze which sources of leakage were exploited by attacks in experiments in prior work.

\xhdr{Data}
For being able to isolate the different sources of leakage, we perform the experiments on synthetic data that is individually crafted for each experiment. Synthetic data also has the advantage that it allows us to directly sample from the possible training distributions \(\mathcal{D}^r\) instead of having to rely on resampling from a finite dataset, which, e.g., typically leads overlap between the training datasets of the different shadow models.
What is common for all experiments except from the one on distributional membership inference (\Secref{sec:experiments_causal}) is that the feature vectors \(X\) are independently sampled from a \(4\)\hyp dimensional normal distribution. In these experiments the label for each feature vector \(X\) is computed by a neural network \(M\) as \(Y=M(X)\). We generated \(M\) as a fully\hyp connected network with one hidden layer with \(32\) neurons and random weights that are sampled independently from the normal distribution \(N(0,1)\). Except for the experiments in \Secref{sec:experiments-distribution_inference-finiteness} regarding the finiteness of the training data, we use \(2048\) training samples for each target and shadow model.
We perform both regression and classification experiments, i.e., we consider adversaries that want to infer the value of a continuous property (\(R=[0,1]\)) and adversaries that want to infer a binary property (\(R=\{0,1\}\)). The value \(r\in R\) is a parameter of the distributions from which we sample \((X,Y)\).

\xhdr{Target model}
The target model in all experiments is a neural network with the same architecture as the model \(M\) that generates the labels. In each experiment we train \(1024\) target models, each with newly sampled data. In the regression case we split the interval \([0,1]\) evenly and train one model for each \(r=k/1023\), where \(k=0,\dots,1023\). In the classification case we train \(512\) models with \(r=0\) and \(512\) models with \(r=1\).

\xhdr{Attacks}
We perform each experiment with the white\hyp box attack developed by Ganju et al. \cite{ganju2018property} and the black\hyp box attack proposed by Zhang et al. \cite{zhang2021leakage}, which are implemented in our new Python library \cite{meynet2022pia}. The white\hyp box attack uses DeepSets \cite{zaheer2017deep} to learn permutation-invariant representations of the models' weights from shadow models, and a classifier or regressor on top of these representations. The black\hyp box attack uses a simple multi\hyp layer perceptron trained on the responses of shadow models to a set of 1024 random queries.
For each attack we train \(2048\) shadow models. From this pool we sample \(100\) times \(1024\) shadow models without replacement to train \(100\) meta\hyp classifiers. We compute the mean accuracy (classification) or mean average error (MAE; regression) of the attack models on the \(1024\) target models, and average these values over the \(100\) attack models. In addition to the means, we show the \(95\%\) confidence intervals w.r.t.\ the \(100\) performance values of the attack models.
An optimal attack that does not take into account the target model would achieve an accuracy of \(0.5\) for classification and an MAE of \(0.25\) for regression.

\subsection{Distribution inference}
\label{sec:experiments-distribution_inference}
\subsubsection{Differences in \(\mathbb{E}[Y|X]\)}
\label{sec:experiments-distribution_inference-y_given_x}

\begin{figure}
    \centering
    \includegraphics[width=0.48\textwidth]{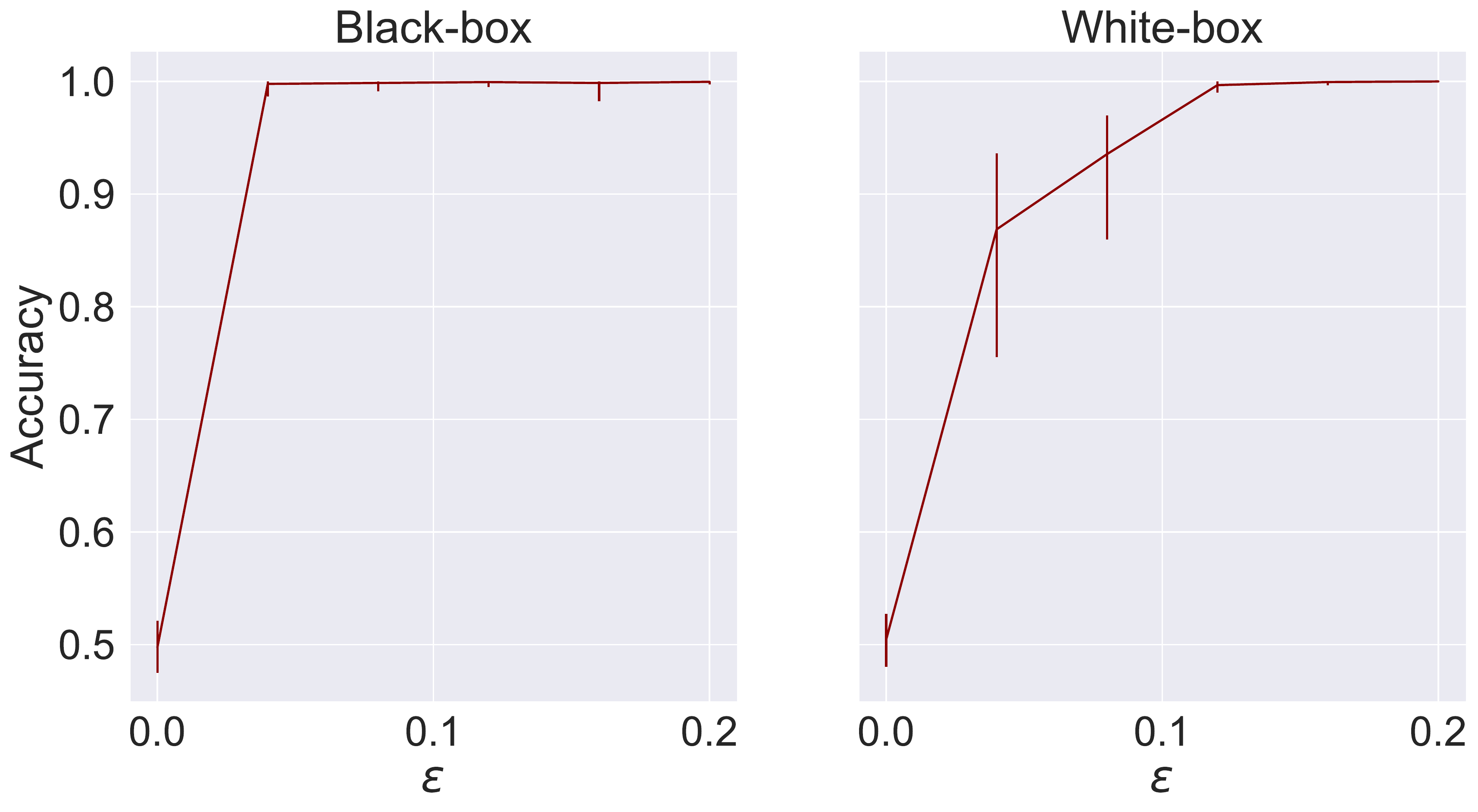}
    \caption{\textbf{First experiment regarding differences in  \(\mathbb{E}[Y|X]\) (\Secref{sec:experiments-distribution_inference-y_given_x}).} \(\varepsilon\) determines how far apart the weights of the two models generating $Y^0$ and $Y^1$ are. Error bars represent \(95\%\) confidence intervals.}
    \label{fig:exp_a}
\end{figure}

We randomly generate a neural network \(M^0\) that computes the value of \(Y^0\) given \(X^0\) as \(Y^0=M^0(X^0)\), as described above. We then generate a neural network $M_1$ that computes \(Y^1=M^1(X^1)\), where for \(M^1\) we take the same architecture, weights and biases as for \(M^0\), but, for a value $\varepsilon \in \mathbb{R}$, add independent noise $N(0, \varepsilon^2)$ to every weight and bias term. We generate the features \(X^b\), for \(b=0,1\), as
$$X \sim N_4(0, 2 I_4),$$
where \(N_4\) denotes the four\hyp dimensional normal distribution and \(I_4\) the four\hyp dimensional identity matrix.
\Figref{fig:exp_a} shows the attack accuracy for different values of \(\varepsilon\). We see that larger differences between the functions that compute \(Y^0\) and \(Y^1\) lead to a higher attack accuracy.

\begin{figure}
    \centering
    \includegraphics[width=0.48\textwidth]{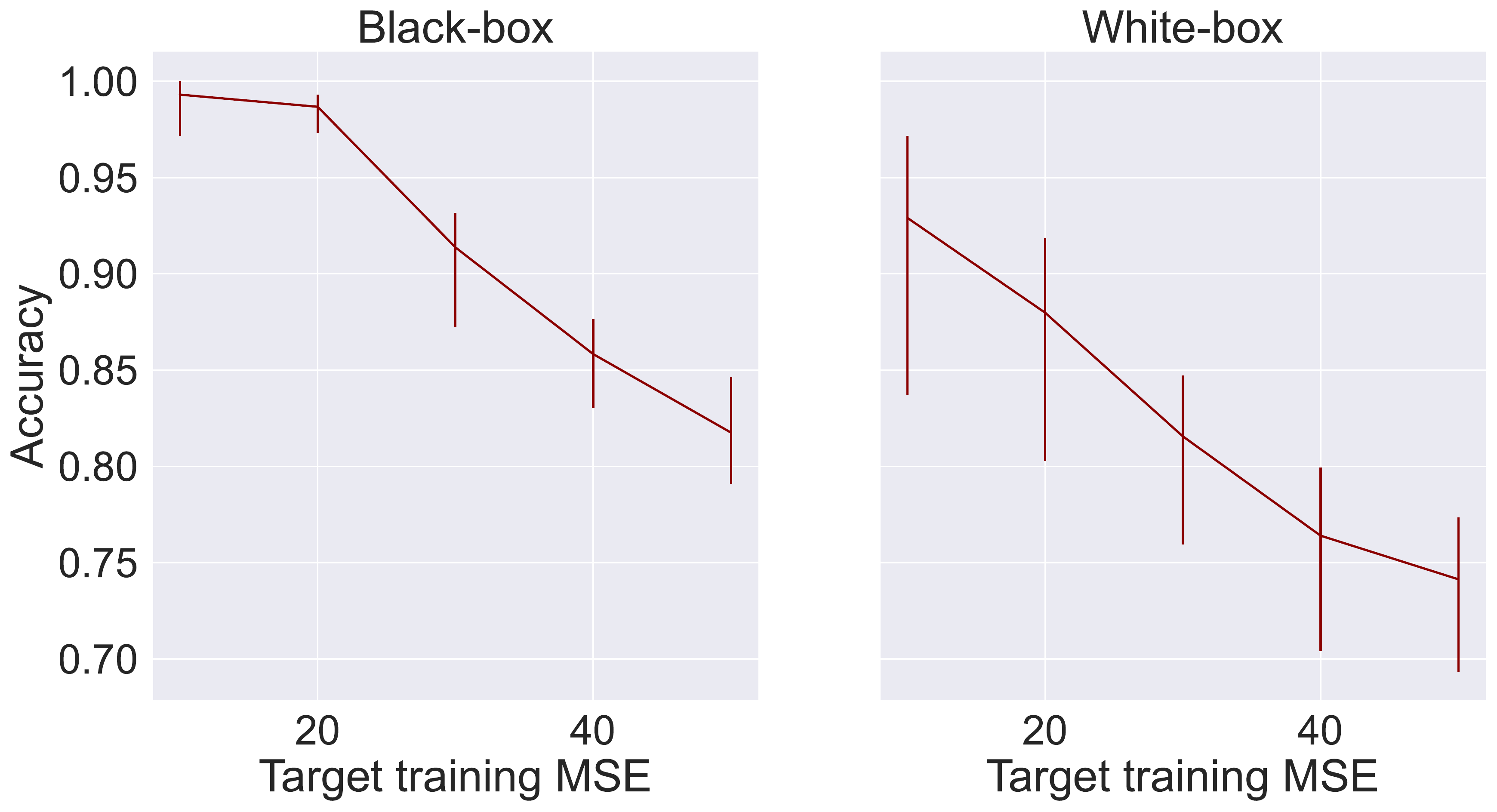}
    \caption{\textbf{Second experiment regarding differences in  \(\mathbb{E}[Y|X]\) (\Secref{sec:experiments-distribution_inference-y_given_x}).} The training of the target model is stopped early after it has reached the predefined training MSE. Error bars represent \(95\%\) confidence intervals.}
    \label{fig:exp_abis}
\end{figure}

We further investigate the influence of the performance of the target model on the attack performance in this setting. To produce models with a predefined performance, we perform early stopping: we only train the model as long as the training MSE lies above the target MSE, and stop training after the first epoch where the training MSE falls below the target MSE. We perform the same early stopping in the training of the shadow models. \Figref{fig:exp_abis} shows the results for different target MSE's at a fixed perturbation level $\varepsilon = 0.05$. As expected, target models that better fit the function \(\mathbb{E}[Y^b|X^b]\) are more susceptible to distribution inference attacks than those that fit it more poorly, in the setting where this function differs between \(\mathcal{D}^0\) and \(\mathcal{D}^1\).

\subsubsection{Wrong inductive bias}
\label{sec:experiments-distribution_inference-inductive_bias}

\begin{figure}
    \centering
    \includegraphics[width=0.48\textwidth]{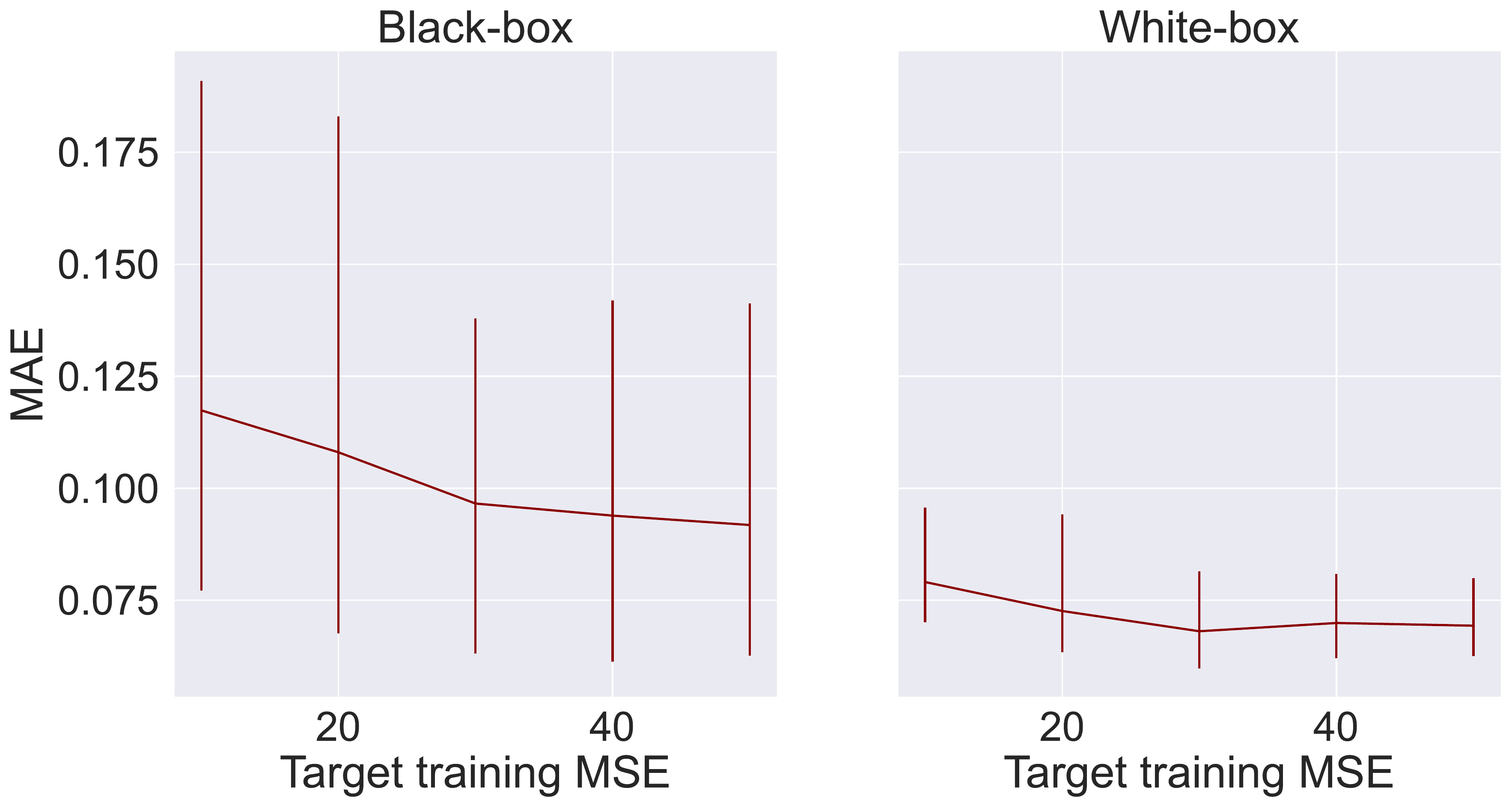}
    \caption{\textbf{Experiment regarding wrong inductive biases (\Secref{sec:experiments-distribution_inference-inductive_bias}).} The training of the target model is stopped early after it has reached the predefined training MSE. Error bars represent \(95\%\) confidence intervals.}
    \label{fig:exp_b}
\end{figure}

With \(\mathbf{1}_4\) denoting the four\hyp dimensional vector of ones, we generate the features \(X^r\) of distribution \(\mathcal{D}^r\) as
$$\textbf{X} \sim N_4((-1 + 2 r)\mathbf{1}_4, 2 I_4),$$
and the labels \(Y^r\) via a random neural network \(M\) as described earlier.
In order to simulate the result of a wrong inductive bias, we perform early stopping in the same way as in the previous experiment (\Secref{sec:experiments-distribution_inference-y_given_x}).
The results in \Figref{fig:exp_b} for the regression case show that the worse the model fits the data, the better the distribution inference attack performs. There is a small uptick in the attack MAE for the white\hyp box case for poorly fitted models, which might be due to a larger variance in the target model early in the training, which makes the learning task for the meta\hyp classifier harder.
We repeat the same experiment for the classification case (\Figref{fig:exp_b_class}), but find no interesting trends since both attacks perform almost perfectly in all cases.

\subsubsection{Finiteness of the training data}
\label{sec:experiments-distribution_inference-finiteness}

\begin{figure}
    \centering
    \includegraphics[width=0.48\textwidth]{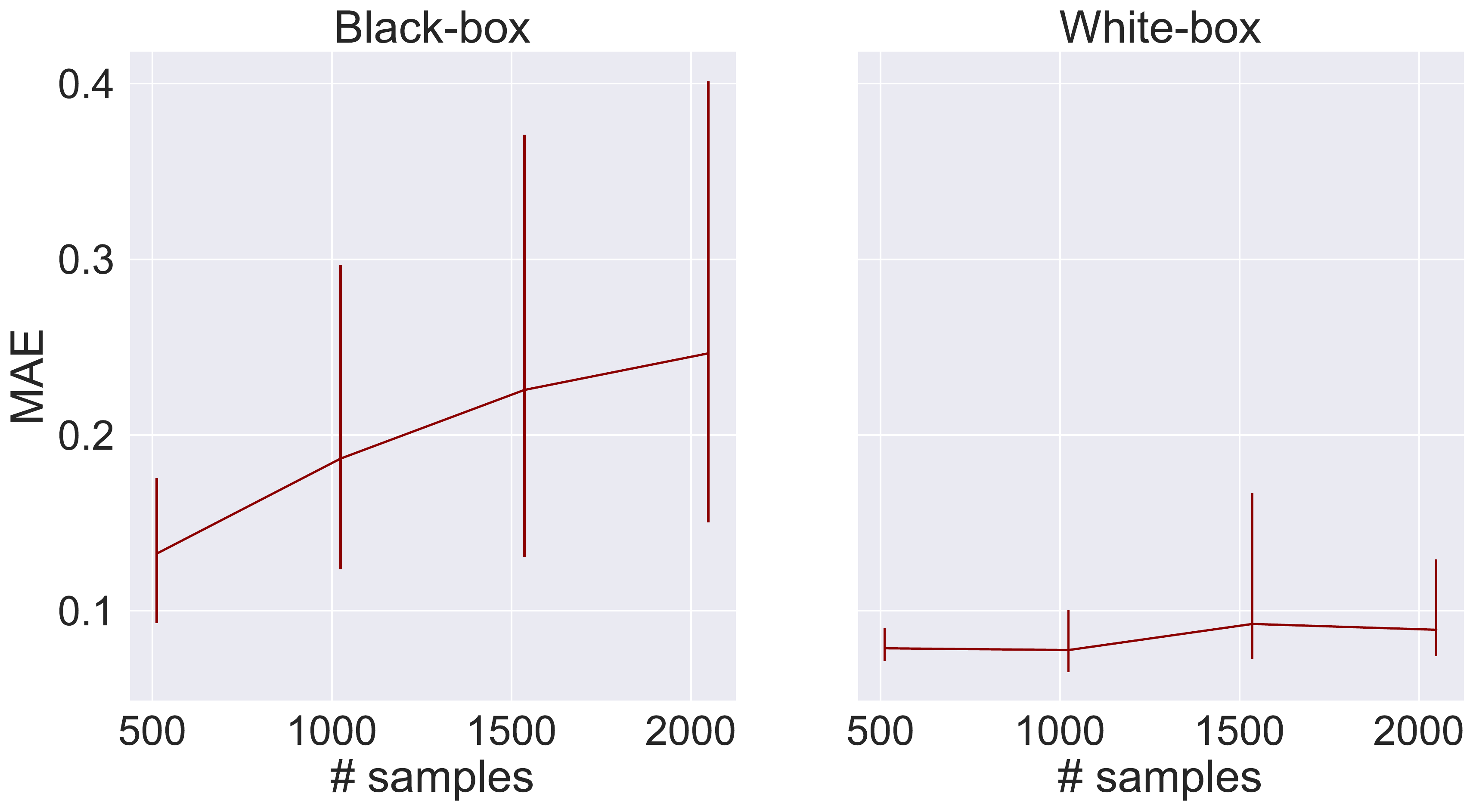}
    \caption{\textbf{Experiment regarding wrong inductive biases (\Secref{sec:experiments-distribution_inference-finiteness}).} The target model gets trained with datasets of different sizes. Error bars represent \(95\%\) confidence intervals.}
    \label{fig:exp_c}
\end{figure}

We generate the data in the same way as in the previous experiment (\Secref{sec:experiments-distribution_inference-y_given_x}). However, instead of using \(2048\) training record for the target and the shadow models, we vary this number between \(512\) and \(2048\). In \Figref{fig:exp_c} we see that a larger training set provides more protection against distribution inference attacks. We perform the same experiment for the classification case (\Figref{fig:exp_c_class}), where we observe the same behavior in the black\hyp box attack, whereas the white\hyp box attack always performs perfectly.

\subsection{Distributional membership inference}
\label{sec:experiments_causal}
In \Secref{sec:defenses-membership_inference} we showed how perfect IRM protects against distributional membership inference attacks if the parties' distributions do not differ in terms of the relationships of the causal parents \(X_C\) of the label \(Y\) with \(Y\). We set up an experiment to see whether such a protection is also possible with existing implementations of the IRM concept. For this, we generate features \(X=(X_1, X_2)\) and a label \(Y\) according to the causal graph \(X_1 \rightarrow Y \rightarrow X_2\). The feature \(X_1\) is a causal parent of \(Y\), whereas \(X_2\) is spuriously correlated with \(Y\). Concretely, we sample data for party \(i\) as follows:
\begin{align*}
    X_1 &\sim \mathcal{N}(0, 1)\\
    Y &\sim X_1 + \mathcal{N}(0, 1)\\
    X_2 &\sim Y + \mathcal{N}(0, 0.5 + i).
\end{align*}

We use four parties with indices \(0, 1, 2, 3\), and sample \(512\) records from each party's distribution. We assume that the adversary wants to determine the presence or absence of the data of party \(3\) in the training data, which means that \(D^0\) consists of the data of all four parties, whereas \(D^1\) only consists of the data of parties \(0\), \(1\) and \(2\). Hence, the correlation between \(X_2\) an \(Y\) is larger in \(\mathcal{D}^1\) than in \(\mathcal{D}^0\) and therefore \(\mathbb{E}[Y^0|X_1^0, X_2^0]\neq \mathbb{E}[Y^1|X_1^1, X_2^1]\). A perfect associational ERM model that takes both \(X_1\) and \(X_2\) into account for its prediction would thus leak distributional information, whereas a perfect causal model would not leak distributional information, since by design \(\mathbb{E}[Y^0|X_1^0] = \mathbb{E}[Y^1|X_1^1]\).

We perform distribution inference attacks on three models that are all fully\hyp connected neural networks with one hidden layer with two neurons:
(1) an ERM model, which takes \(X_1\) and \(X_2\) as input;
(2) an optimal causal ERM model, which only takes \(X_1\) as input;
and (3) an IRM model, where \(\Phi\) is the fully\hyp connected neural network described above with \(X_1\) and \(X_2\) as input and a one\hyp dimensional output, and \(w\) is the constant \(1\), as proposed by Arjovsky et al.\ \cite{arjovsky2019invariant}.

\begin{figure*}
\centering
\begin{subfigure}{.48\textwidth}
    \centering
    \includegraphics[width=\textwidth]{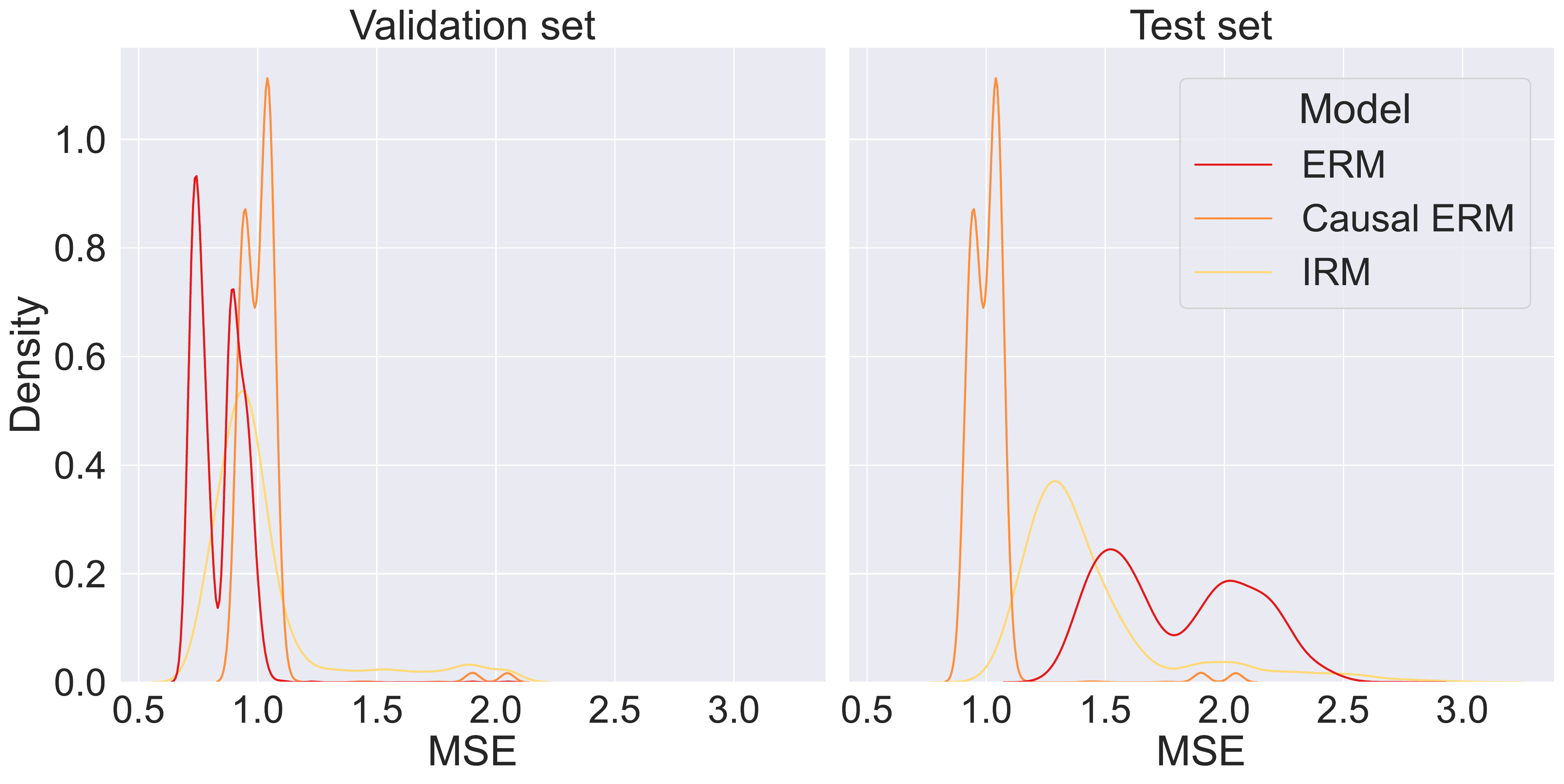}
    \caption{Target model performance.}
    \label{fig:exp_causal_perf}
\end{subfigure}
\begin{subfigure}{.48\textwidth}
    \centering
    \includegraphics[width=\textwidth]{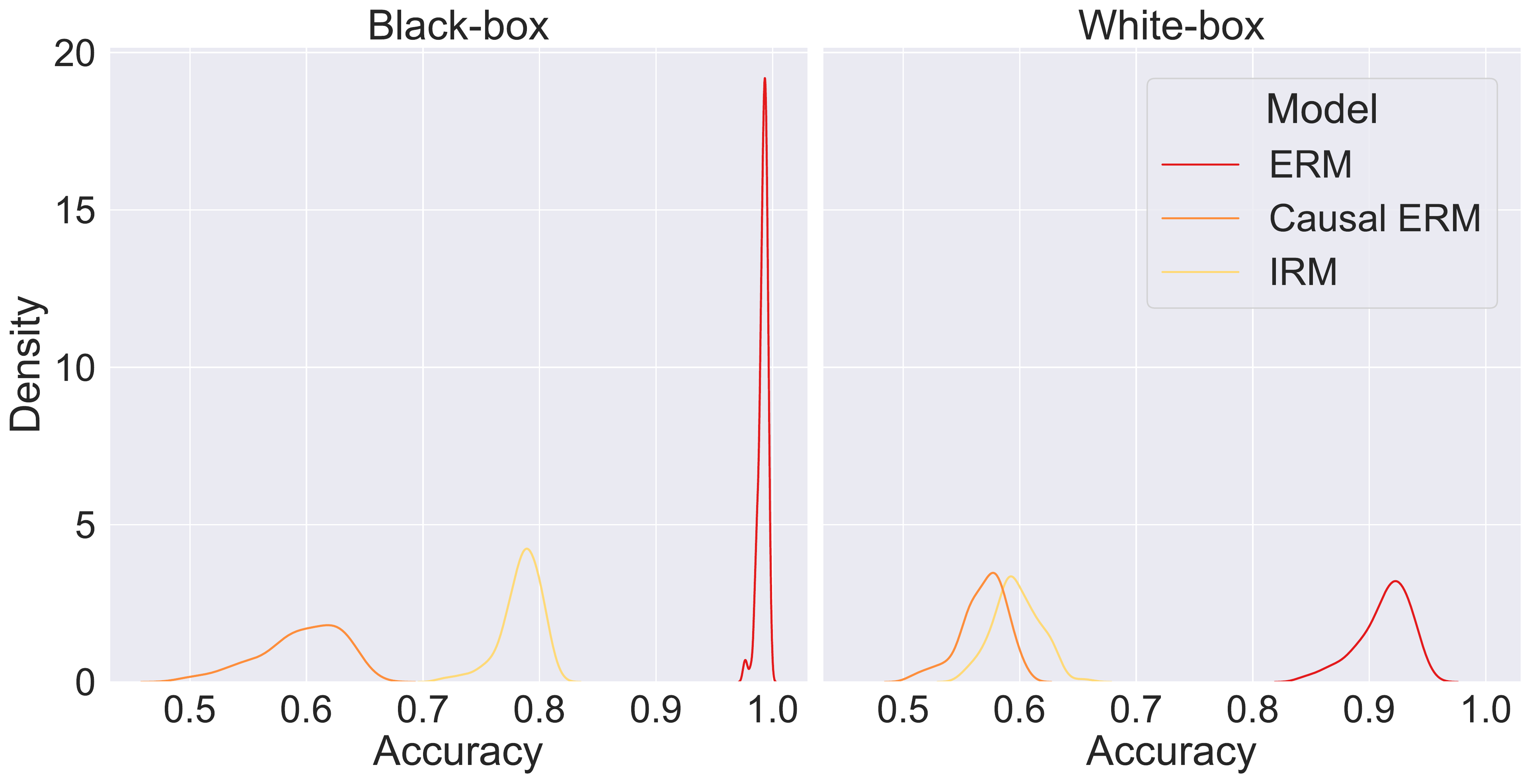}
    \caption{Attack performance.}
    \label{fig:exp_causal_res}
\end{subfigure}
\caption{\textbf{Experiments in the distributional membership inference setting (\Secref{sec:experiments_causal}).} The plots show densities based on (a) 1024 target models of each type and (b) 100 attacks, respectively.}
\end{figure*}

We first assess the in\hyp distribution performance and out\hyp of\hyp distribution generalization capabilities of the different models. For this, we sample new records from the training distributions $\mathcal{D}_0$ and $\mathcal{D}_1$ to create validation sets. We further create test sets from the same distributions, except with the spurious correlation between \(X_2\) and \(Y\) inverted:
\begin{equation*}
    X_2 \sim -Y + \mathcal{N}(0, 0.5 + i).
\end{equation*}
A model that learns to ignore the spurious correlation with \(X_2\) will perform worse on data from the train, but better on data from the test distribution than a model that leverages \(X_2\) for its predictions. And indeed, as shown in \Figref{fig:exp_causal_perf}, the ERM model, which is expected to use both \(X_1\) and \(X_2\) for its predictions, is that model among the three that performs best on the validation set and worst on the test set. For the causal ERM model, which only uses \(X_1\), the situation is reversed. IRM lies between the two, which shows that it achieves a certain level of out\hyp of\hyp distribution generalization, though it is still not as good as the optimal causal model.

Next, we investigate the resistance to distribution inference attacks. \Figref{fig:exp_causal_res} shows the distribution of the attack performance of \(100\) meta\hyp classifiers for each of the three models. As expected, the attacks perform much worse on the optimal causal model than on the ERM model. We can further see that IRM, despite having to learn the causal structure by itself, nearly matches the protection of the optimal causal model against a white\hyp box adversary. In the black\hyp box case the protection is still much better than that of the ERM model, but the gap to the optimal causal model is bigger. We conjecture that this is the result of the higher variance of the IRM model as compared to the causal ERM model, as seen in \Figref{fig:exp_causal_perf}, which could make the learning task for the white\hyp box meta\hyp classifier harder.
Note how neither the optimal causal model nor IRM provide perfect protection, which would correspond to an attack accuracy of \(0.5\). The reason for this is the finiteness of the training data, due to which the models cannot learn the exact relationship between \(X_1\) and \(Y\). However, the dataset \(D^0\) is larger than \(D^1\), and hence the expected deviation from the correct regression parameter is larger when training on \(D^1\), which the attacks can exploit.

\section{Further related work}
\label{sec:related}
\subsection{Distribution inference}
\label{sec:related-distribution_inference}
We give an overview of the existing distribution inference attacks and of the proposed defenses.

\subsubsection{Attacks}
Ateniese et al.\ \cite{ateniese2015hacking} are the first to show the relevance of the problem of distribution inference by performing successful distribution inference attacks. The attacks operate in the white\hyp box setting and consist of a meta\hyp classifier trained on top of shadow models (see \Secref{sec:background-distribution_inference}). The authors show the success of their attacks against support vector machines and hidden Markov models.
Ganju et al.\ \cite{ganju2018property} show that a simple meta\hyp classifier that takes a flattened representation of the target model's weight as input does not work well against complex, fully connected neural networks (FCNNs). They make the observation that FCNNs are invariant w.r.t.\ layer\hyp wise node permutations. Based on this observation they propose two ways to encode this information in the meta\hyp classifier: via permutation\hyp invariant representations or via a representation of the layers of the target model as sets via DeepSets \cite{zaheer2017deep}.
Zhang et al.\ \cite{zhang2021leakage} consider distribution inference in the black\hyp box setting and propose a meta\hyp classifier attack based on shadow models.

Zhang et al.\ also show how their attack can be applied in a secure multi\hyp party ML setting, where multiple parties jointly train an ML model with their local data, and one of the parties wants to gain information about the data distribution of another party that participates in the training.
Mahloujifar et al.\ \cite{mahloujifar2022property} show that data poisoning --- which may happen in multi\hyp party learning as well --- can be used to increase the effectiveness of distribution inference attacks.

Suri and Evans \cite{suri2021formalizing} formalize distribution inference as a cryptographic game. They describe a way to relate the accuracy of an attack to the number of training records that an adversary with sampling access to the training distribution would have to draw to obtain the same amount of information. Further, the authors develop new black\hyp box attacks that are based on the error that the target model makes when performing inference on \(\mathcal{D}^0\) and \(\mathcal{D}^1\). They also extend the white\hyp box attack by Ganju et al.\ \cite{ganju2018property} based on DeepSets to convolutional neural networks. The authors compare the new and existing attacks on different datasets and show that attacks that aim to infer the ratio of a binary target attribute are more successful the further apart the two ratios considered by the adversary are.

Zhou et al.\ \cite{zhou2021property} develop distribution inference attacks against generative adversarial networks (GANs) for black\hyp box settings where the adversary either has access to models outputs generated from random inputs or from inputs chosen by the adversary. Their attacks are the first to perform regression, i.e., they directly output the estimated value of the attacked attribute instead of only distinguishing between two possible values.
Zhou et al.\ further show how distribution inference attacks can be used to enhance membership inference attacks by calibrating the output of the membership inference attacks to the attributes of the target record in combination with the inferred property of the training distribution.

\subsubsection{Defenses}
While the majority of the work on distribution inference focuses on attacks, two papers also discuss possible defenses.

Ganju et al.\ \cite{ganju2018property} propose three defenses and report preliminary results on the first and the third one.
1) They observe that in a neural network with ReLu or LeakyReLu activation functions scaling the weights and bias of a neuron by a factor and scaling the weights connecting it to the next layer by the inverse factor does not change the computed function. A defense that does not affect the model performance can thus consist of performing such scaling and inverse scaling in the target model in a randomized way. However, since this defense does not change the computed function, it does not work against black\hyp box attacks, and might also be overcome by more sophisticated white\hyp box attacks that, e.g., perform the same kind random scalings in the training of the shadow models.
2) Another defense proposed by the authors is to encode additional, arbitrary information in the model parameters to make them look different from those of the shadow models. This defense has similar shortcomings as the first one: it does not work against black\hyp box attacks and might also be overcome by adapted white\hyp box attacks.
3) Finally, the authors propose randomly flipping labels in the training data. However, this directly impacts the model performance and, as the authors themselves write, is therefore unlikely to be implemented in real\hyp world applications.

Zhou et al.\ \cite{zhou2021property} propose two defenses without investigating them empirically.
1) The first defense is specific to GANs. Instead of releasing all samples generated by a GAN, the model owner could release only a specific subset of the generated samples that is less susceptible to the attacks. However, this would mean changing the output distribution of the GAN, which would make it less closely resemble the training distribution.
2) The second defense can also be applied to other models. The authors propose to modify the training dataset w.r.t.\ the target property prior to training, either by removing or by adding records so that the ratio of a targeted binary property has a fixed predetermined value, e.g., \(0.5\), no matter the original ratio. This may mean expensive acquisition of new data or reducing the size of the dataset, potentially significantly.
Both of these defenses also assume that the defender knows the property that the attacker wants to attack.

As we can see, all proposed defenses come with significant drawbacks and lack thorough experimental evaluation. We hope that the insights from this paper can support the development of new, practical defenses. Our Python library that implements state\hyp of\hyp the\hyp art distribution inference attacks will be particularly useful for evaluating such defenses.

\subsection{Membership inference}
In this section we describe previous settings for membership inference and position our definition of distributional membership inference in this space of definitions.

ML models have been shown to be susceptible to record\hyp level membership inference attacks \cite{shokri2017membership}, that is, an adversary with access to the model may be able to determine with good accuracy whether a particular record was part of the training dataset or not. Multiple attacks have been developed \cite{shokri2017membership,salem2019ml,yeom2018privacy,long2017towards}, some of them relying on the idea of shadow models that is also used by many distribution inference attacks.

A natural defense against record\hyp level membership inference attacks is differential privacy (DP) \cite{dwork2006calibrating}, which guarantees that the output distribution of a randomized function does not change too much when changing one record in the database that it is invoked on.
A DP mechanism has a privacy parameter \(\varepsilon\), and, in the case of approximate DP, additionally a privacy parameter \(\delta\) \cite{dwork2006our}. Yeom et al.\ \cite{yeom2018privacy} formalize record\hyp level membership inference as an adversarial game and show that when the training algorithm fulfills \(\varepsilon\)\hyp DP, the adversary's advantage can be bounded by a function of \(\varepsilon\).

To achieve DP for an ML model, one can for example randomly perturb the learning objective or the parameters of the model \cite{chaudhuri2011differentially}. For neural networks trained via SGD, instead a perturbation of the gradients is used \cite{abadi2016deep,mironov2019renyi}.
Such perturbations usually lead to a loss in utility.

Classic DP assumes that each individual that contributed to the database contributed exactly one record. Under this assumption, preventing an adversary from inferring the membership of one record is equivalent to preventing the adversary from inferring whether one individual has contributed to the database. However, in many cases there might be individuals who have contributed multiple records. E.g., each visit of the same patient to a hospital might lead to a new record in the hospital's database, or similarly each query by the same user to a search engine might lead to a new record in the provider's database. DP can be extended to these settings by increasing the standard deviation of the noise added in the function computation by a factor of the maximal number of records contributed by one individual \cite{dwork2008theory}, which will decrease utility. Furthermore, such a bound on the maximal number of contributed records may not even be known. Most variants of DP cannot handle this last setting, whereas it poses no problem for distributional membership inference.

There exist numerous variants of DP \cite{desfontaines2020sok,jayaraman2019evaluating}. Some of them are related to our setting of distributional membership inference insofar as they also assume an adversary that only has distributional information about (some of) the potential records in the database instead of knowing their precise values.

Hall et al.\ \cite{hall_random_2013} propose random differential privacy, where the dataset is assumed to consist of i.i.d.\ samples from the same distribution, and they consider an adversary that wants to distinguish between two datasets that are identical except for one record that is drawn independently in the datasets.
Triastcyn and Faltings \cite{triastcyn2020bayesian} define Bayesian differential privacy, where they only consider a single record in the data as random, and assume that this record is present in one dataset but absent in the other.
Duan's \cite{duan2009privacy} work is closer to ours in that it assumes that no random noise is added in the computation of the function whose result is supposed to fulfill a privacy guarantee. Duan assumes that the adversary only has access to some of the records in the dataset, but that there are \(n\) records for which the adversary only knows the distribution that they were sampled from. Duan shows that due to the central limit theorem in this setting sum queries fulfill DP for large enough \(n\) even without adding external noise.

In recent work, Suri et al.\ \cite{suri2022subject} study subject membership inference where, like us, they assume an adversary with only distributional information about the different individual's datasets that make up the training data. They develop attacks for a federated learning setting where each subject may have contributed to the data of multiple parties that participate in the training.

\subsection{Causal learning}
\label{sec:related-causal_learning}
The connection between causal learning and privacy has been explored in prior work.
Tople et al.~\cite{tople2020alleviating} investigate the resilience of causal models to membership inference attacks. They show that such models provide better protection than associational models and achieve stronger DP guarantees for the same amount of noise.
Similarly, Francis et al.\ \cite{francis2021towards} experimentally show that models trained via a federated learning (FL) version of IRM are more resilient to membership inference attacks than associational models trained via FL.

\section{Conclusion}
This paper provides the first in\hyp depth study of the reasons for the leakage of information about the training distribution of ML models.
We have identified the three sources of leakage via theoretical analysis --- memorization of information about \(\mathbb{E}[Y|X]\), a wrong inductive bias, and the finiteness of the training data ---, and corresponding defense strategies.
Our experiments show that state\hyp of\hyp the\hyp art distribution inference attacks exploit all of these sources, and that our defense strategies can reduce the attack performance.
We have further introduced a new, general definition of distribution inference that allows for handling a broader class of adversaries.
Beyond this generalization, we have introduced and analyzed the new, specialized attack scenario of distributional membership inference and shown how causal learning methods can provide protection against such attacks.
We hope that our uncovering of the mechanisms behind distributional leakage, along with our Python library for distribution inference attacks, can help in the development of new attacks and especially defenses.


\bibliographystyle{abbrv}
\bibliography{references}


\appendix
\subsection{Additional experiments}
\label{app:further_experiments}
We repeat the experiments from \Secref{sec:experiments-distribution_inference-inductive_bias} and \Secref{sec:experiments-distribution_inference-finiteness} with classification instead of regression attacks. Since this is a much easier task, the attacks perform very well in any case, making the experiments less suitable to investigate the influence of a wrong inductive bias and of finite datasets on the attack performance.

\begin{figure}[hbt]
    \centering
    \includegraphics[width=0.48\textwidth]{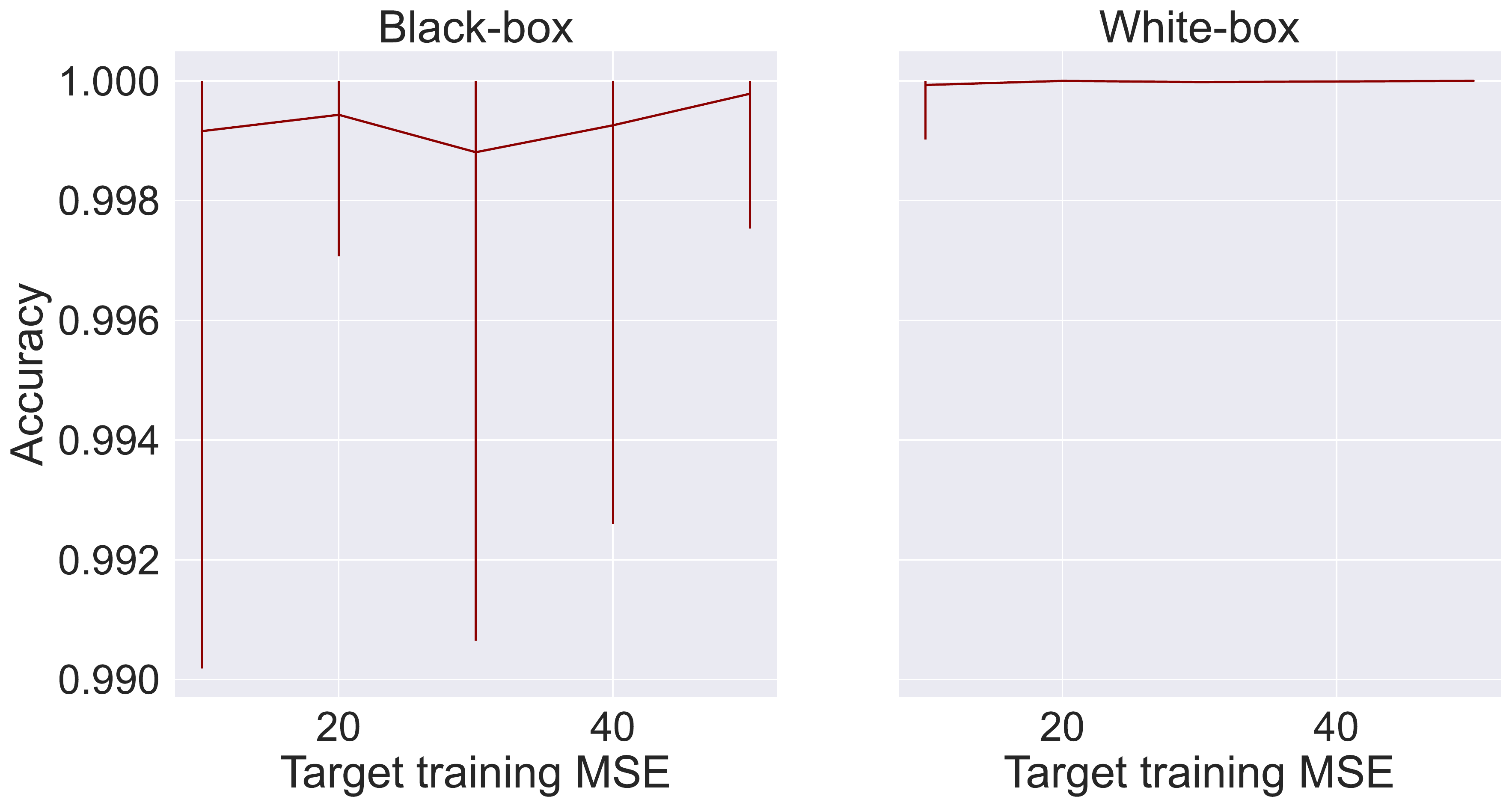}
    \caption{\textbf{Experiment regarding wrong inductive biases (\Secref{sec:experiments-distribution_inference-inductive_bias}); classification attack.} The training of the target model is stopped early after it has reached the predefined training MSE. Error bars represent \(95\%\) confidence intervals.}
    \label{fig:exp_b_class}
\end{figure}

\begin{figure}[hbt]
    \centering
    \includegraphics[width=0.48\textwidth]{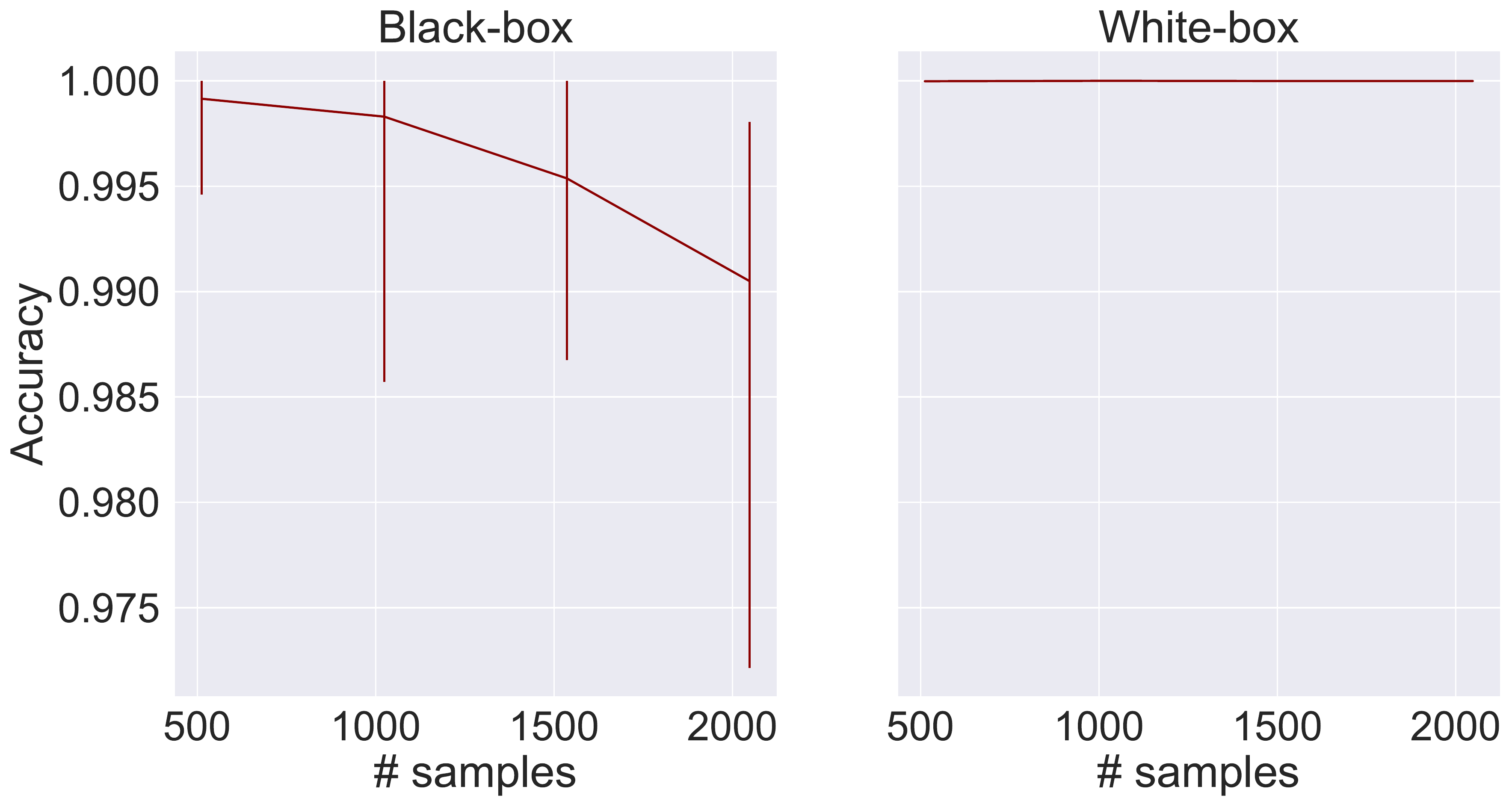}
    \caption{\textbf{Experiment regarding wrong inductive biases (\Secref{sec:experiments-distribution_inference-finiteness}); classification attack.} The target model gets trained with datasets of different sizes. Error bars represent \(95\%\) confidence intervals.}
    \label{fig:exp_c_class}
\end{figure}

\subsection{Analyzing subsampling-based experiments}
\label{app:subsampling}
In \Secref{sec:experiments-distribution_inference} we have presented the results of controlled experiments that each focus on one of the three sources of training distribution leakage. Prior to our paper, these different sources had not been identified and therefore experiments, which were typically used for comparing the performance of different attacks, did not deliberately include or exclude some of the sources of leakage. If an experiment, however, does not include one source of leakage and a particular attack's strength is picking up exactly on this source of leakage, then the attack performance might be underestimated.
For real datasets, the inductive bias of a model is almost never completely correct, and the amount of training data is always finite. Therefore these two sources of leakage are almost always present in experiments on real data. However, it is not always the case that \(\mathbb{E}(Y^0|X^0) \neq \mathbb{E}(Y^1|X^1)\). We will hence investigate in which experiments in the prior work on distribution inference this source of leakage is present and in which not.

In most of these experiments \(\mathcal{D}^0\) and \(\mathcal{D}^1\) are generated by taking one dataset and conditionally subsampling from it to generate two datasets that have different marginal distributions of a target attribute \(T\) whose distribution the adversary wants to learn. Sampling from \(\mathcal{D}^0\) and \(\mathcal{D}^1\) is then defined as sampling from these two datasets. \(T\) can be one of the features \(X\) (e.g., the sex of an individual in a census dataset \cite{ganju2018property}), but does not have to be (e.g., the age of the displayed person in an image dataset \cite{ganju2018property} whose features are pixel colors).
Typically \(T\) is a binary attribute --- which we, too, assume in this section for simplicity --- and \(\mathcal{D}^0\) and \(\mathcal{D}^1\) therefore differ in the probability that the target attribute takes the value \(0\) (or \(1\)). Analogous to \(X^b\) and \(Y^b\), let \(T^b\) be the random variable corresponding to the target attribute with marginal distribution according to \(\mathcal{D}^b\). The data generation process has the properties
\begin{equation*}
    \Pr(Y^0|X^0,T^0) = \Pr(Y^1|X^1,T^1)
\end{equation*}
and
\begin{equation*}
    \Pr(T^0=0) \neq \Pr(T^1=0).
\end{equation*}
For fixed \(x\), \(y\), we have
\begin{align*}
    &\Pr(Y^b=y,X^b=x)\\
    &= \sum_{t\in \{0,1\}} \Pr(Y^b=y,X^b=x| T^b = t) \Pr(T^b = t).
\end{align*}
We distinguish between the case where \(T^b\) is one of the features \(X^b\) and the case where it is not one of the features.

\xhdr{\(T\) is a feature} Suppose that \(X^b\) is a feature vector of length \(m\) and that \(T^b\) is one of the features in \(X^b\), w.l.o.g.\ the last feature \(X^b_m\): \(X^b=(X^b_{<m}, T^b)\). Let \(x\) be an instance of \(X^b\) with \(x_m=0\). Then \(\Pr(X^b=x|T^b=1)=0\) and \(\Pr(Y^b=y,X^b=x| T^b=1)=0\). We thus have
\begin{align*}
    &\Pr(Y^0=y|X^0=x)\\
    &= \frac{\sum_{t\in\{0,1\}} \Pr(Y^0=y,X^0=x| T^0=t) \Pr(T^0=t)}{\sum_{t\in\{0,1\}} \Pr(X^0=x|T^0=t)\Pr(T^0=t)}\\
    &= \frac{\Pr(Y^0=y,X^0=x| T^0=0) \Pr(T^0=0)}{\Pr(X^0=x|T^0=0)\Pr(T^0=0)}\\
    &= \frac{\Pr(Y^0=y,X^0=x| T^0=0)}{\Pr(X^0=x|T^0=0)}\\
    &= \frac{\Pr(Y^1=y,X^1=x| T^1=0)}{\Pr(X^1=x|T^1=0)}\\
    &= \Pr(Y^1=y|X^1=x).
\end{align*}
Hence, in the case where \(T\) is one of the features, it does not hold that \(\mathbb{E}(Y^0|X^0) \neq \mathbb{E}(Y^1|X^1)\), and therefore in such experiments this potential source of leakage cannot be exploited.

\xhdr{\(T\) is not a feature} This is not necessarily the case if \(T^b\) is not one of the features \(X^b\). Consider the case where \(X^b\) and \(Y^b\) are binary scalar random variables. For \(p_0 \neq p_1\), let
\begin{alignat*}{2}
    \Pr(Y^b=0|X^b=0, T^b=0) &= p_0,\\
    \Pr(Y^b=0|X^b=0, T^b=1) &= p_1.
\end{alignat*}
We have
\begin{align*}
    &\Pr(Y^b=0|X^b=0)\\
    &= \sum_{t\in\{0,1\}}\Pr(Y^b=0|X^b=0, T^b=t)\Pr(T^b=t)\\
    &= p_0\Pr(T^b=0) + p_1\Pr(T^b=1).
\end{align*}
Assume that \(\Pr(T^0=0) = p_0\) and \(\Pr(T^1=0) = p_0\). Then \(\Pr(Y^0=0|X^0=0)\neq\Pr(Y^1=0|X^1=0)\) and therefore \(\mathbb{E}(Y^0|X^0)\neq\mathbb{E}(Y^1|X^1)\), because
\begin{align*}
    &\Pr(Y^0=0|X^0=0) - \Pr(Y^1=0|X^1=0)\\
    &= [p_0*p_0 + p_1(1-p_0)] - [p_0*p_1 + p_1(1-p_1)]\\
    &= p_0^2 -2p_0 p_1 + p_1^2\\
    &= (p_0-p_1)^2\\
    &> 0,
\end{align*}
since by assumption \(p_0\neq p_1\).

\end{document}